\documentclass[conference]{IEEEtran}
\usepackage{cite}
\usepackage{amsmath,amssymb,amsfonts}
\usepackage{algorithmic}
\usepackage{graphicx}
\usepackage{textcomp}
\usepackage{xcolor}
\usepackage[hyphens]{url}
\usepackage{listings}
\usepackage{mathptmx} 
\usepackage{color,soul}
\usepackage{fancyhdr}
\usepackage[normalem]{ulem}
\usepackage{amsmath,amssymb,amsfonts}
\usepackage{algorithmic}
\usepackage{graphicx}
\usepackage{textcomp}
\usepackage{xcolor}
\usepackage{fancyhdr}
\usepackage{tabularx}
\usepackage{comment}
\usepackage{balance}

\def\BibTeX{{\rm B\kern-.05em{\sc i\kern-.025em b}\kern-.08em
    T\kern-.1667em\lower.7ex\hbox{E}\kern-.125emX}}
\newcommand{\proj}{Dalorex}
\newcommand{\device}{TSU}

\title{Dalorex: A Data-Local Program Execution and Architecture for Memory-bound Applications}

\author{\IEEEauthorblockN{Marcelo Orenes-Vera,
Esin Tureci, David Wentzlaff and
Margaret Martonosi}
\IEEEauthorblockA{
Princeton University\\
Princeton, New Jersey, USA\\
Email: \{movera, esin.tureci, wentzlaf, mrm\}@princeton.edu }}

\begin{document}
\maketitle
\pagestyle{plain}

\newcommand{\subf}[2]{%
  {\small\begin{tabular}[t]{@{}c@{}}{\tiny}
  #1\\#2
  \end{tabular}}%
}

\definecolor{codegreen}{rgb}{0,0.6,0}
\definecolor{codegray}{rgb}{0.5,0.5,0.5}
\definecolor{codepurple}{rgb}{0.58,0,0.82}
\definecolor{backcolour}{rgb}{0.97,0.97,0.96}
\lstdefinestyle{mystyle}{
    backgroundcolor=\color{backcolour},   
    commentstyle=\color{codegreen},
    keywordstyle=\color{blue},
    numberstyle=\tiny\color{codegray},
    stringstyle=\color{codepurple},
    basicstyle=\ttfamily\scriptsize,
    breakatwhitespace=false,         
    breaklines=true,                 
    captionpos=b,                    
    keepspaces=true,                 
    numbersep=5pt,                  
    showspaces=false,                
    showstringspaces=false,
    showtabs=false,                  
    tabsize=2,
    morekeywords={task, task1, task2, task3, task4, dist, edge_idx, ptr, edge_values, OQ1,OQ2,OQ3,OQ4,IQ1,IQ2,IQ3,IQ4}
}

\begin{abstract}

Applications with low data reuse and frequent irregular memory accesses, such as graph or sparse linear algebra workloads, fail to scale well due to memory bottlenecks and poor core utilization.
While prior work with prefetching, decoupling, or pipelining can mitigate memory latency and improve core utilization, memory bottlenecks persist due to limited off-chip bandwidth.
Approaches doing processing in-memory (PIM) with Hybrid Memory Cube (HMC) overcome bandwidth limitations but fail to achieve high core utilization due to poor task scheduling and synchronization overheads.
Moreover, the high memory-per-core ratio available with HMC limits strong scaling.

We introduce Dalorex, a hardware-software co-design that achieves high parallelism and energy efficiency, demonstrating strong scaling with \textgreater{16,000} cores when processing graph and sparse linear algebra workloads.
Over the prior work in PIM, both using 256 cores, Dalorex improves performance and energy consumption by two orders of magnitude through
(1) a tile-based distributed-memory architecture where each processing tile holds an equal amount of data, and all memory operations are local;
(2) a task-based parallel programming model where tasks are executed by the processing unit that is co-located with the target data;
(3) a network design optimized for irregular traffic, where all communication is one-way, and messages do not contain routing metadata;
(4) novel traffic-aware task scheduling hardware that maintains high core utilization;
and (5) a data-placement strategy that improves work balance.

This work proposes architectural and software innovations to provide the greatest scalability to date for running graph algorithms while still being programmable for other domains.

\end{abstract}
\begin{IEEEkeywords}
Distributed, scalable, parallel, data-local, near-memory, architecture, sparse, graph, NoC, network, bandwidth.
\end{IEEEkeywords}

\vspace{-3mm}
\section{Introduction}

System designs are increasingly exploiting heterogeneous, accelerator-rich designs to scale performance due to the slowing of Moore’s Law~\cite{moorelaw} and the end of Dennard Scaling~\cite{goldenage,42y}.
While compute-bound workloads thrive in an accelerator-rich environment, memory-bound workloads, such as graph algorithms and sparse linear algebra, present the following \textbf{challenges}:
(a)~a low compute-per-data ratio, with little and unpredictable data reuse;
(b)~frequent fine-grained irregular memory accesses that make memory hierarchies inefficient;
(c)~atomic accesses, synchronization, and inherent load imbalance resulting in poor utilization of computing resources.

Recent work proposed using accelerators that pipeline the different stages of graph processing in hardware~\cite{graphicionado,graphpulse,ozdal,chronos,polygraph, fifer}.
Other works have used general-purpose cores with decoupling and prefetching to mitigate memory latency~\cite{desc,graphattack,maple,pipette,prodigy}, and coalescing to reduce atomic update serialization~\cite{phi}, but failed to avoid costly data movements and are eventually bottlenecked by memory bandwidth.
The processing-in-memory (PIM) proposals enjoy a higher memory bandwidth by processing data near DRAM~\cite{tesseract, 2019graphq, 2018graphp}.
However, their memory integration constrains the \textbf{storage-per-core} ratio, limiting the level of parallelism---as we demonstrate in this paper. 
In addition, their parallelization suffers from load imbalance and synchronization overheads.

\begin{figure}[t]
\centering  
\vspace{-2.5mm}
\includegraphics[width=\columnwidth]{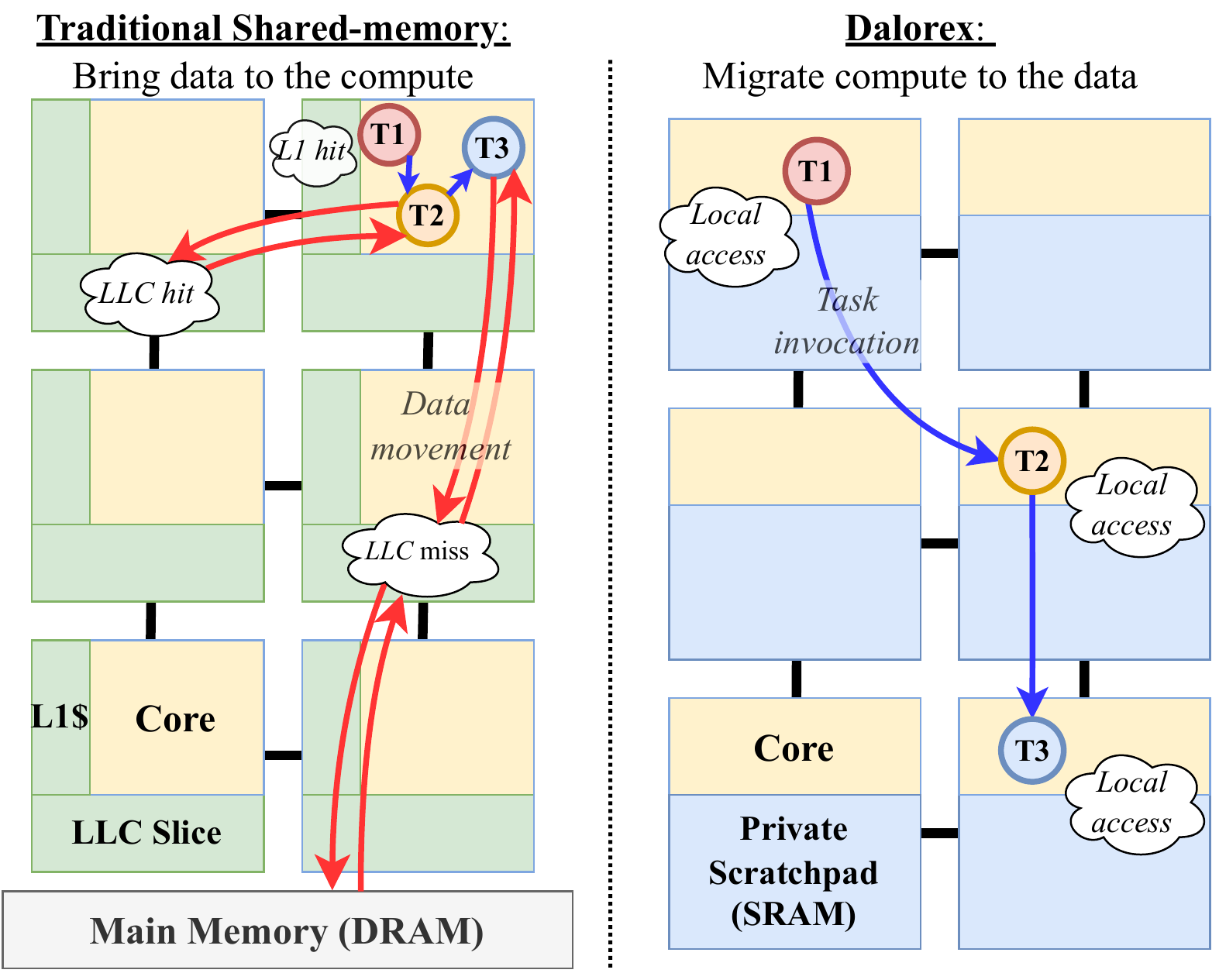}
\vspace{-5mm}
\caption{
Program execution of three sequential graph-processing steps in a cache hierarchy (left), and \proj{} (right).
Instead of moving data with little reuse, \proj{} invokes tasks where the data is local---reducing movement.
}
\vspace{-4.5mm}
\label{fig:approach}
\end{figure}

\textbf{Opportunities:}
We started our work by examining what features are necessary to execute graph workloads in a scalable manner.
We observed that to maximize throughput (edges processed per second), \textit{the solution should}:
(1)~minimize data movement, which is the performance bottleneck and the most significant energy cost;
(2)~exploit the spatio-temporal parallelism of operations to improve work balance and resource utilization;
(3)~avoid serializing features such as global synchronization barriers and read-modify-write atomic operations; 
and (4)~avoid frontier redundancy and data staleness to minimize the number of edges explored (work efficiency).

\begin{figure}[t]
\centering  
\includegraphics[width=\columnwidth]{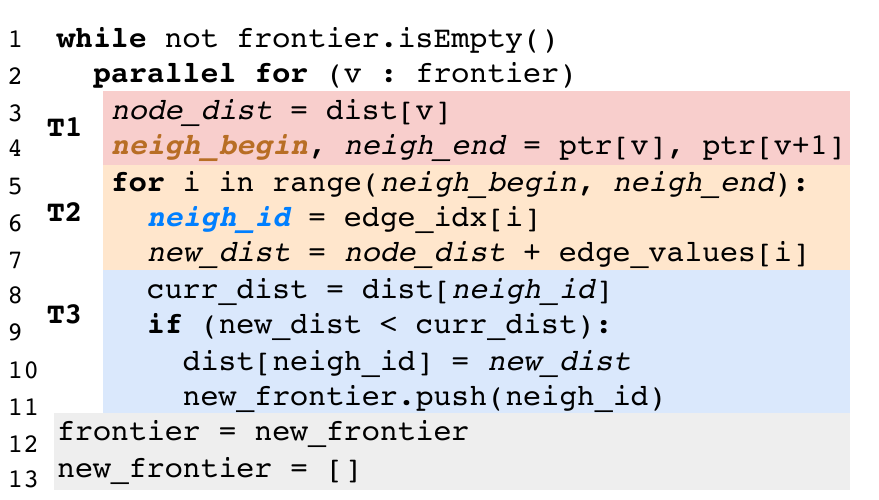}
\vspace{-7mm}
\caption{Pseudocode of the Single Source Shortest Path (SSSP) algorithm, and how it is split into \proj{} tasks based on indirect memory accesses. The colored variables determine the tile at which the next task is executed.
}
\vspace{-4mm}
\label{fig:sssp_motivation}
\end{figure}

\textbf{Our Approach:}
We present \proj{}, a highly scalable and energy-efficient hardware-software co-design that exploits these opportunities by optimizing across the computing stack without sacrificing ISA programmability.
Our architecture is a 2D array of homogeneous tiles, each containing an SRAM scratchpad, a thin in-order core (with no cache) called Processing Unit (PU), a task scheduling unit feeding the PU, and a router connected to the network-on-chip (NoC).
Our programming model enables further parallelism by splitting the sequential code inside a parallel-loop iteration into tasks at each pointer indirection (Fig.~\ref{fig:sssp_motivation}).
Tasks execute at the tile containing the data to be operated on.
Correct program order is guaranteed because new tasks are spawned by the parent task by placing their invocation parameters into the NoC.
Since the spawned tasks are independent and execute in any order, \proj{} achieves huge synchronization-free spatio-temporal parallelism.
We designed the task scheduling unit (\device{}) with a closed-loop feedback system to achieve work efficiency and high utilization, as this varies with task flow order in this decentralized model.
\device{} also coordinates PUs and routers to make task invocations non-blocking and non-interrupting.

\proj{} distributes all data arrays in equal chunks across tiles using low-order index bits.
In this way, \proj{} achieves both good workload and NoC traffic balance, despite the irregular access patterns of sparse workloads.
As a result of this data partitioning, each tile owns a set of nodes (with no copies of data), and all operations are atomic by design.
Because of this, \proj{} can also have a decentralized frontier, such that one can remove the global barrier between epochs of graph algorithms.
In addition, the data-local execution of tasks removes the round-trip time and energy of accessing memory, as well as the serialization due to access contention.

Fig.~\ref{fig:approach} illustrates how the distributed memory architecture of \proj{} (right panel) minimizes data movement compared to architectures with hierarchical shared-memory structures (left panel).
In \proj{}, each piece of data is only accessed by one PU.
Instead of bringing data to the PUs, they send task-invoking messages to the tile containing the data to be processed next.
\proj{} exploits the inherent pointer indirection in sparse data formats to route a task-invocation message.
The tile's PU executes the task corresponding to the message received.
In architectures with a traditional memory hierarchy, sparse applications result in overwhelming data movement: data-reuse distance is highly irregular, and thus, cache thrashing leads to more than 50\% of all memory accesses missing in the cache levels and going to main memory~\cite{graphattack}.
Shared-memory architectures also carry the overhead of cache coherence, virtual memory, and atomic operations.



\textbf{The technical contributions of this paper are}:

    
\begin{itemize}
    \item A data-local execution model where each processing tile holds an equal amount of data---operating on local data makes all updates atomic and minimizes data movement.
    \item A programming model that unleashes parallelism and improves work balance by splitting the code into tasks---based on data location---while preserving program order.
    \item A tiled distributed-memory architecture connected by a NoC optimized for irregular communication (headerless task routing based on the index of the array they access).
    \item A hardware unit (\device{}) that removes task-invocation overheads and schedules tasks to maximize core utilization and work efficiency by sensing the network traffic.

\end{itemize}

\textbf{We evaluate \proj{} and demonstrate that:}
\begin{itemize}
    \item Our architecture and data layout improve \textit{performance} by 6.2$\times$ over the best ISA-programmable prior work~\cite{tesseract}.
    On top of that, our task invocation scheme improves by 4.7$\times$, and our uniform data placement and traffic-aware scheduling, 4.4$\times$ more. 
    Finally, removing the barriers and upgrading the NoC provides an extra 1.8$\times$, compounding a 221$\times$ geomean improvement across four key graph applications, using equal processor count (256).
    \item Regarding \textit{energy}, the improvements of using SRAM ($16\times$), our parallelization and data placement (5.2$\times$), and \device{} (3.9$\times$) compound 325$\times$, in geomean.
    \item \proj{} achieves strong scaling with over 16,000 processing tiles, not being limited by memory bandwidth.
    \item Our data distribution and parallelization do not require pre-processing to improve work balance across tiles; \proj{} remains agnostic to the dataset characteristics.
\end{itemize}

\section{Background and Motivation}\label{sec:background}

\proj{} is designed to accelerate applications that are memory-bound due to irregular access patterns caused by pointer indirection. 
Although this paper focuses on graph algorithms, \proj{} is applicable to other domains such as sparse linear algebra.
We demonstrate this by evaluating sparse matrix-vector multiplication.

\subsection{Graph Algorithms and their Memory Access Patterns}

Graphs are represented using adjacency matrices where rows/columns represent vertices and values, weighted edges.
Since columns contain mostly zero values (most vertices have few connections), these matrices are stored in formats like Compressed-Sparse-Row (CSR) using four arrays.
The non-zero elements are accessed via pointer indirection.

Fig.~\ref{fig:sssp_motivation} shows the sequential code for Single Source Shortest Path (SSSP) and colors the code sections that would be split into \proj{} tasks.
In a regular memory hierarchy, the accesses to neighbor vertex data in the innermost loop (\texttt{line 8}) result in many cache misses and thus, costly accesses to DRAM~\cite{graphattack}.
Moreover, the source vertex data (\texttt{lines 3 and 4}) and the neighbor index (\texttt{line 6}) are also accessed indirectly, and the utility of the cache depends on the number of neighbors and the workload distribution.
In \proj{}, the code is split at each level of pointer indirection, leading to a series of tasks. For example, \texttt{T1} accesses arrays \textbf{dist} and \textbf{ptr} (tuple of size $\#vertices$), while \texttt{T2} accesses arrays \textbf{edge\_idx} and \textbf{edge\_values} (tuple of size $\#edges)$, and \texttt{T3} accesses \textbf{dist}.

\begin{figure}[t]
\centering  
\vspace{-1mm}
\includegraphics[width=\columnwidth]{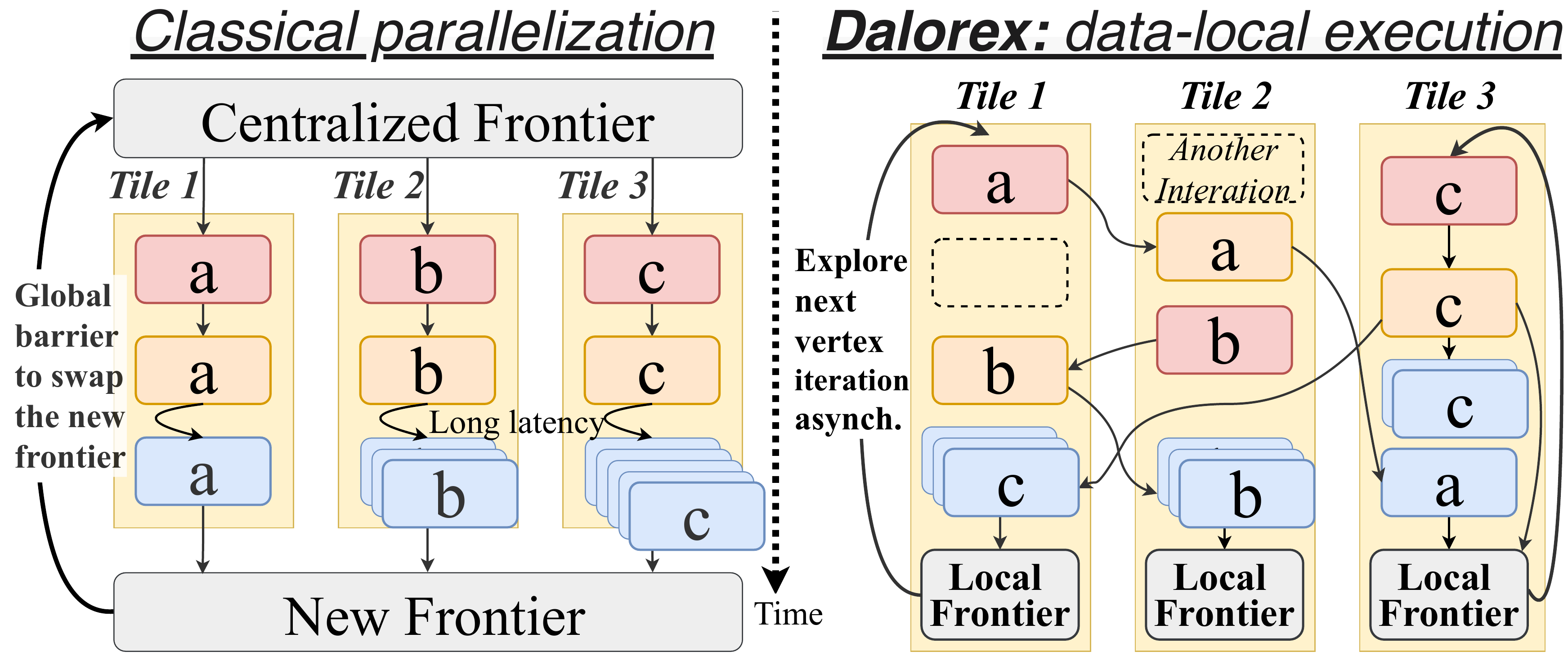}
\vspace{-4mm}
\caption{Program order and synchronization for Bulk Synchronous Parallel (BSP), left, versus \proj{}.
BSP parallelization leads to work imbalance as Task2 (orange) may generate several Task3 (blue).
Columns show the tasks that are executed in each tile (arrows depict program order).
Colors indicate task type based on the code of Fig.~\ref{fig:sssp_motivation}.
Letters represent different vertex iterations.
Dotted boxes depict that tasks from other iterations may interleave.
} 
\vspace{-3mm}
\label{fig:program_flow}
\end{figure}

As Fig.~\ref{fig:program_flow} illustrates, \proj{} allows \textit{spatial interleavings within vertex iterations} while preserving program order through sequential invocation of the tasks.
The classical time-wise interleavings of parallel-loop iterations are also allowed.
Finally, instead of using a centralized frontier, \proj{} uses local frontiers, which removes the synchronization overheads of frontier insertions and global barriers.
As a result, our novel programming model maximizes program parallelization.

\vspace{-1mm}
\subsection{Algorithmic variants}

There are two modes of processing graph data: pulling data for a vertex from its neighbors or pushing data to its neighbors~\cite{push_pull}.
While pull-based algorithms tend to require higher memory communication, push-based algorithms often require atomic operations.
When atomic operations are mitigated, push-based algorithms have the advantage of reduced communication overhead.
A hybrid version, direction-optimized \textit{BFS} \cite{direction-optimized-BFS}, and its variants can offer faster convergence.
However, they incur a storage overhead and need heuristics, as they may access either the source or the destination of an edge.
Although pull-based algorithms can be executed on the \proj{} architecture, we focus on push-based algorithms due to their reduced communication and work efficiency since \proj{} eliminates the need for atomic operations.


\vspace{-1mm}
\subsection{Prior Work}
We describe pipelining and data-movement-reducing techniques and lay out what aspects limit their scalability.

\subsubsection{Hardware Techniques}
Recent works have used decoupling to overlap data fetch and computation by running ahead in the loop iterations to bring the data asynchronously~\cite{pipette, graphattack,prodigy,maple}.
To accomplish this, they perform program slicing on each software thread, creating a software pipeline effect. 
Others have proposed accelerators to perform the graph search as a hardware pipeline~\cite{graphicionado,graphpulse,ozdal,chronos, fifer}.
Polygraph~\cite{polygraph} generalized prior accelerator designs to perform any of their algorithmic variants and optimize work efficiency based on dataset characteristics, and Fifer~\cite{fifer} offers a dynamic temporal pipelining to achieve load-balancing.
While effective for hiding latency and increasing load-balancing, these approaches remain highly energy inefficient due to excessive data movement and are ultimately limited by DRAM bandwidth.

To increase memory bandwidth and reduce data movement, Tesseract~\cite{tesseract} proposes in-memory graph processing by introducing cores into the logic layer of a 3D Hybrid Memory Cube (HMC)~\cite{hmc,hmc_cc}.
Tesseract executes remote calls at the cores located near the data, similar to the execution-migration literature~\cite{emmachine,livia}.
However, their performance is limited because:
(a)~Their vertex-based data distribution causes load imbalance since the highly-variable vertex degree in graphs causes a different workload per core;
(b)~Tesseract remote calls are interrupting, incurring 50-cycle penalties, and GraphQ's solution to overcoming this employs barriers for batch communication~\cite{2019graphq}, causing high synchronization overheads;
(c)~HMC-based architectures are constrained in the number of cores per cube (a core per vault).
In this work, as we explore the limits of graph parallelization, we demonstrate that the energy-optimal storage size per tile is in the single-digit megabyte range, much smaller than what HMC offers.


\subsubsection{Software Techniques}
Software solutions to accelerating graph applications include optimizing data placement, work efficiency, and parallelization schemes based on the target hardware system~\cite{graphit}.
For distributed systems, GiraphUC~\cite{barrierless} puts forward a barrierless model that reduces message staleness and removes global synchronization.
However, GiraphUC is not optimized for data locality and has high communication costs.
Pregel~\cite{pregel} does parallelize tasks such that each iteration is executed where the data is local, but the data is distributed in a vertex-centric manner, resulting in inherent load-balancing problems and communication overheads.

\subsection{Manycore Architectures \& Memory Bandwidth}\label{sec:background_hmc}

Large-scale parallel processing can use many small units.
From Systolic arrays~\cite{systolic_78,systolic_93,tpu_google} and streaming architectures~\cite{raw,operand_networks,streaming_algorithms}, to modern manycores~\cite{piton, manticore, esperanto, celerity}, supplying irregularly-accessed data remains challenging.

Recently, some industry products have utilized large amounts of SRAM to achieve high on-chip bandwidth, and so, higher performance~\cite{cerebras,graphcore}. 
Aside from energy consumption, there are architectural advantages to using a scratchpad memory per core, e.g., low latency, dedicated access, and scalable memory bandwidth (more cores means more memory ports).

\section{\proj{}: A Hardware-software Co-design}\label{sec:approach}

\proj{} minimizes data movement and maximizes resource utilization with:
~(1) a data distribution that allows for \textit{only} local memory operations;
~(2) a programming model that allows for intra-loop spatial and inter-loop timewise interleavings;
~(3) a homogeneous tile-based architecture that optimizes irregular data-access patterns, where each tile includes a local memory, a router, and a processing unit.
Program execution is orchestrated by the task scheduling unit (\device{}).

\subsection{Data distribution}\label{sec:data_distibution}
Graphs and sparse matrices are often stored in formats like CSR using four data arrays.
In \proj{}, these arrays are divided equally across all tiles and stored in their private memories, making each tile responsible for operations only on its local data.
For example, the \texttt{edge\_values} array has as many elements as edges ($E$) in a graph. 
This array is split as in Listing~\ref{lst:sssp_code} so that each of the $T$ tiles has \texttt{EDGES\_PER\_CHUNK} ($E/T$) adjacent elements, e.g., the first tile contains elements from $0$ to \texttt{EDGES\_PER\_CHUNK-1}.

Ours is the first work that distributes an adjacency matrix in this manner; the usual approach is to do a 2D distribution of the matrix~\cite{2d_graph_partitioning}, where each computing element gets a rectangular subset of the matrix to compute.
Although designed to minimize communication, 2D distribution presents some challenges: a subset of a sparse matrix is hyper-sparse (making row or column sparse formats storage-inefficient), and the resulting chunks do not have an equal number of non-zeros (different storage needs).
Because \proj{} has all memory equally distributed across tiles, it improves workload balance.

\subsection{Programming model}\label{sec:prog_model}
The BSP model parallelizes graph algorithms either at the outer loop (processing vertices in the frontier) or the inner loop (processing the neighbors of the frontier vertices) since these can be processed in any order.
We posit that if the iterations of the inner loop are performed in program order, the location of the execution can be altered.
\proj{} preserves the instruction order within an iteration since subsequent tasks are invoked by the parent task at completion.
Since only one tile has access to each data chunk, coherence is not an issue.

Adapting a graph kernel to \proj{} involves splitting the inner-loop iteration into multiple tasks at each pointer indirection.
As we have seen in Fig.~\ref{fig:sssp_motivation}, this results in three tasks for SSSP, where each task produces the array index to be accessed by the next task.
Additionally, graph algorithms require a fourth task to explore the frontier vertices (Listing~\ref{lst:sssp_code}).
Every tile contains the same code and can perform any task.

\textit{From the program execution timeline perspective}, after a tile performs a task, it sends the output of the task (i.e., the input for the next task) to the tile containing the data to be operated next, thus preserving sequential order.

\textit{From the point of view of an individual tile}, inputs for any task may arrive in any order into their corresponding task-specific queues.
The execution order of different tasks is determined by the \device{}---described in Section~\ref{sec:hardware_sec}.

Application programmers would not program \proj{} directly.
Instead, DSLs such as TACO~\cite{taco} (sparse algebra) or GraphIt~\cite{graphit} (graph analytics) could invoke our kernel library.

\begin{lstlisting}[language=Python, caption={Pseudo-code of the SSSP algorithm adapted to use the \proj{} programming model. This programming model would be embedded in a high-level language via an API. The programmer or compiler would fill tasks' code.}, label=lst:sssp_code, style=mystyle,float]
## These constants are filled when the program is loaded
param NODES_PER_CHUNK 
param EDGES_PER_CHUNK
## Local chunk of the dataset arrays
var dist[NODES_PER_CHUNK] 
var ptr[NODES_PER_CHUNK]
var edge_idx[EDGES_PER_CHUNK]
var edge_values[EDGES_PER_CHUNK]

const FRONTIER_LEN = NODES_PER_CHUNK/32
const OQT2 = 1024
## Bitmap frontier and memory-stored variables
var frontier[FRONTIER_LEN] = [0,...,0]
var blocks_in_frontier = 0
var t1_new_vertex = True
var neighbor_begin = 0;
 
##Configure network channels between tasks and their queues 
CQ1 = channel(q_len=128, target=T2, encode=EDGES_PER_CHUNK)
CQ2 = channel(q_len=OQT2,target=T3, encode=NODES_PER_CHUNK)

## Declaring a task requires the length of its IQ and
## whether its parameters are loaded before the invocation
task T1 [32] ():
   #T1 params aren't pre-loaded. We read from the IQ of T1
   vertex_id = peek(IQ1.head)
   if (t1_new_vertex):
       neighbor_begin = ptr[vertex_id]
   neighbor_end = ptr[vertex_id+1]
   while (!CQ1.full && (neighbor_begin < neighbor_end)):
     ##Split msg if range crosses chunk limits or > OQT2
      tile = (neighbor_begin/EDGES_PER_CHUNK) + 1;
      partial_end = min(neighbor_end, tile*EDGES_PER_CHUNK)
      partial_end = min(partial_end, neighbor_begin + OQT2)
      CQ1 = neighbor_begin ## global idx for tile address
      CQ1 = (partial_end % NODES_PER_CHUNK) ## local idx
      CQ1 = dist[vertex_id]
     neighbor_begin = partial_end
   ## We pop vertex_id if the whole range was pushed to CQ1
   t1_new_vertex = (neighbor_begin == partial_end)
   if (t1_new_vertex):
       pop(IQ1.head)

# Task parameters are loaded by TSU before the task begins
task T2 [128] (neighbor_begin, neighbor_end, vertex_dist):
    for i in range(neighbor_begin,neighbor_end):
    ##Writing to a Channel Queue sends data to the network
        CQ2 = edge_idx[i]
        CQ2 = edge_values[i] + vertex_dist
 
task T3 [2048] (neigh_id, new_dist):
    curr_dist = dist[neigh_id]
    if (new_dist < curr_dist):
        dist[neigh_id] = new_dist

     	## Insert vertex into Local Frontier
     	blk_id = neigh_id >> 5;
        blk_bits = frontier[blk_id];
        frontier[blk_id] = mask_in_bit(blk_bits, neigh_id)
        if (blk_bits == 0): ##only add newly active blocks
            blocks_in_frontier++
            IQ4 = blk_id

# T4 re-explores the local frontier queue        
task T4 [FRONTIER_LEN] ():
    frontier_block = peek(IQ4)
    while(blocks_in_frontier > 0 && !IQ1.full):
        blk_bits = frontier[frontier_block]
        block_base = frontier_block << 5;
        while(blk_bits > 0 && !IQ1.full):
            idx = search_msb(blk_bits)
            blk_bits = mask_out_bit(blk_bits, idx)
            vertex = block_base + idx
            IQ1 = vertex
        # If all the no pending vertices in the block we
        # remove the block from the frontier queue
        if (blk_bits == 0):
            pop(IQ4)
            blocks_in_frontier--
            frontier_block = peek(IQ4)
\end{lstlisting}

\subsection{Program Flow and Synchronization}

\proj{} programs do not have a \textit{main}.
Instead, tiles await the task parameters to arrive in the corresponding input queue, and PUs are invoked by \device{} to process them.
A task invokes the next task by placing the task parameters into an output queue (OQ).
An OQ can be either another task's input queue (IQ) if it operates over data residing in the same tile or a channel queue (CQ), which puts a message into the network.

Listing~\ref{lst:sssp_code} contains the code of the \proj{}-adapted SSSP kernel.
To start running SSSP, only the tile containing the root of the graph search receives a message to invoke the first task.
\textbf{\texttt{T1}} obtains the range array indices that contain the neighbors of \texttt{vertex\_id}.
If this range crosses the border of a chunk, a separate message is sent to each tile with the corresponding \textit{begin} and \textit{end} indices.
Similarly, the range is split if the length is bigger than a constant \texttt{OQT2}, which is set to guarantee that \texttt{T2} can execute without exceeding the capacity of \texttt{CQ2}.
However, \texttt{T1} does not have this guarantee and needs to check explicitly that \texttt{CQ1} does not overflow.
If \texttt{CQ1} fills before sending all the messages for \texttt{vertex\_id} range, a flag is set, and \texttt{T1} ends early after updating the begin index.
The next time \texttt{T1} is invoked it continues operating on the same \texttt{vertex\_id} since it was not popped from IQ1.
Note that \texttt{vertex\_id} was explicitly loaded with \textit{peek}, as opposed to \texttt{T2} and \texttt{T3}, where the task parameters are implicitly popped from their IQs by \device{} (Section~\ref{sec:hardware_sec}).
\textbf{\texttt{T2}} calculates the new distances to all the neighbors of \texttt{vertex\_id} from the root using their edge values and sends this value to the owner of \texttt{T3} data.
\textbf{\texttt{T3}} checks whether the distance of \texttt{neigh\_id} from the root is smaller than the previously stored value. 
If so, \texttt{neigh\_id} needs to be inserted in the frontier.

\proj{} does not use a barrier after each epoch to explore the new global frontier.
Instead, each tile has a local frontier to allow for a continuous flow of tasks.
The \textbf{\textit{local frontier}} is a bitmap that accumulates the updates to the vertices that a tile owns.
\textbf{\texttt{T4}} is responsible for re-exploring the local frontier.
To avoid iterating over every 32-vertex block of the bitmap when \texttt{T4} is invoked, \texttt{T3} pushes the ID of a new block to be explored (\texttt{blk\_id}) into \texttt{IQ4}.
\texttt{T4} then reads from \texttt{IQ4} and pushes the vertices into \texttt{IQ1} so that they are processed again.


\textbf{\textit{Remote invocation:}}
When placing the parameters of the next task into a CQ, the first flit is the index of the distributed array to be accessed so that the message is routed to the tile containing the data for that index (details in Section~\ref{sec:hardware_sec}).
Sending tasks to other tiles is akin to the non-blocking remote function calls employed by Tesseract~\cite{tesseract}. 
Unlike Tesseract, \proj{} task invocations are non-interrupting when received.

\textbf{\textit{Termination:}}
The program ends when all tiles are idle.
This is determined by aggregating a hierarchical, staged, idle signal from all the tiles (co-located with the clock and reset signals).
Similar to a loosely-coupled accelerator~\cite{piccolboni_hpec17,carloni:dac15}, the host gets an interrupt when the global idle signal is set to notify that the work is completed.

\textbf{\textit{Synchronization:}}
Although we strive to avoid synchronization between graph epochs, \proj{} also supports global synchronization by reusing the idle signal.
To have synchronization per epoch, \texttt{T3} should not push new frontier vertices into the IQ of \texttt{T4} and simply add them to the bitmap frontier. 
When all vertices are processed, the host detects that the chip is idle, and it sends a message to all tiles to trigger \texttt{T4} and explore the new epoch.
We characterize performance with and without epoch synchronization in Section~\ref{sec:results}. 

\textbf{\textit{Host:}}
The host is a commodity CPU that arranges the load of the program binary and the dataset from disk to the chip.
The program, composed of tasks and memory-mapped software configurations, is distributed as a broadcast and is identical for all the allocated tiles.
The data is distributed such that every tile receives an equal-sized chunk of each array.

\proj{} does not use virtual memory, although the host processor can still use virtual addressing within its memory.
This avoids the overheads of address translation, which are exacerbated in graph applications due to irregular memory accesses  \cite{graphs_thp}.
Another advantage of having private, uncontested access to memory is not needing to deal with memory coherence or consistency issues.

\subsection{Graph size vs. Chip size}
\proj{} uses a homogeneous 2D layout of tiles.
Because the size of a tile's local memory is determined at fabrication time, the aggregated chip storage scales linearly with the computing capacity (see Section~\ref{sec:scalability} for the storage-to-compute ratio).
\proj{} could be deployed on a $151 mm^2$ chip, with $16\times16$ tiles and 512MB memory, or at wafer-scale, with $256\times256$ tiles and 128GB memory.

\textit{Processing Larger Graphs:}
Building on previous work on distributed graph processing~\cite{giraph++,metis}, we propose using graph partitioning to split larger graphs into multiple parts, where each part is computed on a separate \proj{} chip.
The aggregated storage on-chip determines the maximum size of a graph partition.
Smaller graphs can also be computed concurrently in rectangular subsets of tiles within a chip, uncontested, as they do not share hardware resources.
The tiles that are not allocated to run programs are switched off.


Selecting the number of tiles a program runs on has a lower bound: it should run a subset of \proj{} that is large enough to fit the dataset into the aggregated memory capacity. 
It can also use a larger number of tiles to parallelize even further.
Section~\ref{sec:results} shows that \proj{} scales close to linearly until the parallelization limits are hit when a tile handles less than a thousand vertices.

\begin{figure*}[t]
\centering
\vspace{-4mm}
\includegraphics[width=\textwidth]{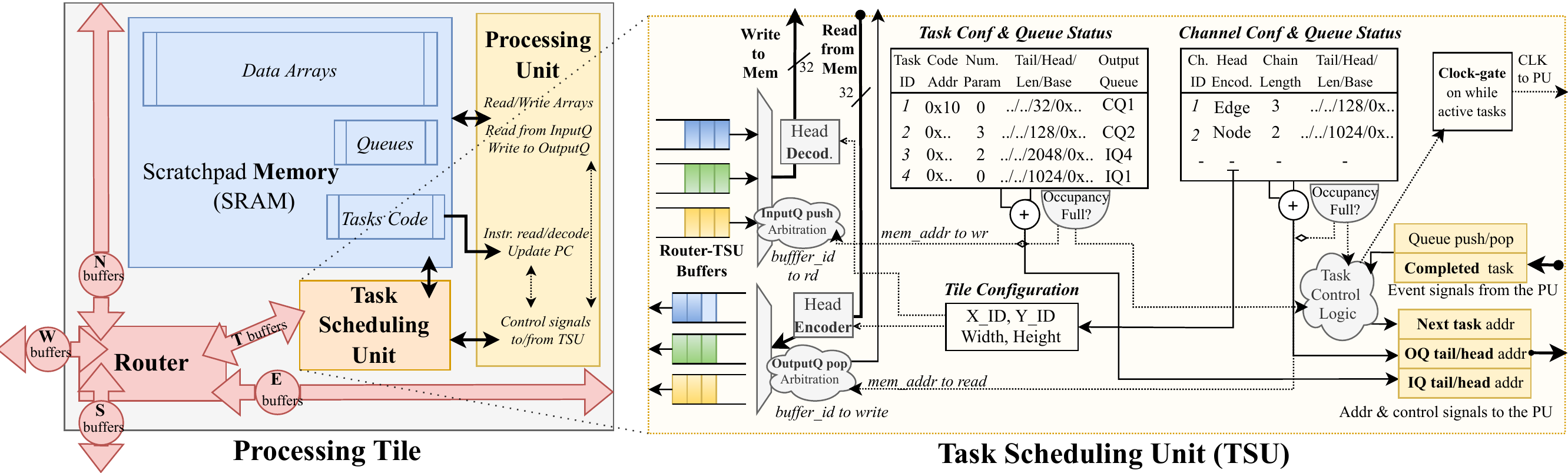}
\vspace{-6mm}
\caption{
Organization of a processing tile.
\device{} feeds the PU the next task to execute based on the occupancy of the task queues.
The router connects \device{} with the network.
\device{} uses SRAM to write incoming network data (push) to the IQs and reads (pop) outgoing data from the channel queues (CQs).
}
\vspace{-4mm}
\label{fig:tile}
\end{figure*}

\subsection{\proj{} Hardware}\label{sec:hardware_sec}

Fig.~\ref{fig:tile} shows the structure of a tile in the \proj{} architecture.
The Processing Unit (PU) is a very power- and area-efficient unit that resembles a single-issue in-order core but without a memory management unit or L1 cache, as the data is accessed directly from the scratchpad memory.
The tile area is dominated by the scratchpad SRAM, which contains the data arrays, the code, and the input/output queue entries.
The queues are implemented as circular FIFOs using the scratchpad.
Queue sizes are configured at runtime based on the number of entries specified next to the task declaration (see Listing~\ref{lst:sssp_code}).
A queue entry can be either 32 or 64 bits, depending on the chip's target total memory size.
(A 32-bit \proj{} can process graphs of up to $2^{32}$ edges.)

The \textbf{Task Scheduling Unit} (\device{}) is the key hardware unit of \proj{}'s hardware-software co-design.
It contains the task configurations and scheduling policy and handles the queues' head and tail pointers.
The \device{} has a read-write port into the scratchpad for pushing data from the router buffers to the input queue (IQ).

\textbf{Queue-specific registers:}
The tail and head pointers of queues are exposed to software through dedicated registers.
This allows the PU to read or write from its queues with a register operation, avoiding address calculation.
A read from this register results in a load from the scratchpad using the corresponding queue's head pointer provided by the \device{}. 
This read also triggers the update of the head pointer in hardware at the \textit{Task Queue Status} table.

\textbf{Scheduling:}
\device{} is responsible for invoking tasks based on the status of the queues, resulting in a closed-loop feedback system.
Tasks cannot block; they execute from beginning to end.
\device{} may only invoke a task if its IQ is not empty and its OQ has sufficient free entries.
When all IQs are empty, \device{} disables the clock of the PU to save power.
\device{} needs to arbitrate when two or more tasks have non-empty IQs.

\textbf{Priority:}
The queues' occupancy acts as a sensor for the \device{} to decide which task to prioritize.
A task has three priority modes based on queue occupancy: high priority if its IQ is nearly full, medium priority if its OQ is nearly empty, and low priority otherwise.
When two or more tasks have high/medium priority, the one with a larger queue size takes precedence.
Testing several static priorities and round-robin schemes, we found that the occupancy-based priority worked best because:
(1)~A large source of network contention is end-point back-pressure, so preventing full IQs eases contention;
(2)~Reaching high resource utilization relies on tiles giving each other work, so keeping OQs non-empty is beneficial.
Using an occupancy-based priority order results in a feedback loop for task scheduling achieving high core utilization and low network contention.
These heuristics are micro-coded as event-condition-action rules based on the reaction time to prioritize a task and prevent the IQ from getting full.


\textbf{Channel queue:}
During program execution, when a task outputs the parameters for the next task to execute on another tile, a channel queue places the data into the network, which is delivered to its destination.

\textbf{Network channel:} It connects a channel queue with a task's IQ.
Messages can be composed of several flits, each being a parameter of the task to be called.
Network communication is composed of flits traveling in different logical channels that share the same network-on-chip (NoC).
In our experiments, a flit has the same size as a queue entry, which is the width of the PU's ALU and the memory addresses (32 bits).

\textbf{Router:}
It has bi-directional ports to north, south, east, west, and towards the tile's \device{}.
Routing is determined by the router based on the data in the first flit of a message, which we call the \textit{head}.
Since the head flit always contains a dataset-array index, and these arrays are statically distributed across the chip, this index is used to obtain the destination tile.
The \device{}'s channel table contains the sizes of the local chunks and the number of parameters (flits) of each message type.
The head encoder uses that information to calculate the destination tile and the local index (modulo the chunk size).
The encoder also uses the width of the \proj{} grid to obtain the X/Y coordinates.
Depending on the width and height of the chip, the upper bits of the head flit encode the destination tile ($log_2(width)+log_2(height)$).
Routers compare incoming head flits with their local X/Y tile ID to determine where to route it next.
If routed to the \device{}, the head decoder removes the head flit's tile-index bits before pushing it to the IQ.

The payload-based routing saves network traffic as it does not use metadata.
The length of messages at each channel is known, and its flits are always routed back to back since a route (from input to output port) opens with the first flit and closes after the corresponding number of flits has left the router.
Interleaving flits from two messages going to the same output port on the same channel is not allowed.
The router can route to different output ports simultaneously. However, it arbitrates by doing a round-robin between the messages from inputs ports that want to route to the same output port.

\textbf{Channels buffers:}
In addition to identifying the task type, communicating tasks in different channels prevents deadlocks.
Although channels might contend to use the NoC, the routers contain buffers per channel so that a clogged channel does not block others.
Each router has a pool of buffer slots per outbound direction shared between the channels.
The size of this pool is a tapeout parameter, but the number of buffer slots per channel is software-configurable, as are the sizes of the input/output queues. 


\subsection{Communication Network}\label{sec:network_sec}
Since \proj{}'s programming model uses task invocations that do not return a value,
communication is one way only, resembling a software pipeline.
Therefore, the communication latency between the sender and receiver tiles does not contribute to the execution time if the pipeline is full, i.e., task invocations are continuous. 
However, the throughput would suffer if bubbles were formed due to network contention.

Graphs vertices have variable degrees.
Some vertices---often referred to as hot vertices---have much higher vertex-degree than others.
The channel buffers and queues mitigate the communication peaks caused by hot vertices and help keep the pipeline effect.
\emph{\proj{} uses low-order bits of indices to distribute data randomly, so the number of hot vertices per tile is relatively uniform.}
Should the graph be sorted by vertex degree, we build the global CSR so that consecutive vertices fall into different tiles.
This uniformity avoids excessive end-point contention at the \device{} with messages from the router.

We observed that the network topology could be another source of contention.
\proj{} uses a 2D torus to avoid the contention towards the center we observed on a 2D mesh.
Our 2D torus is a wormhole network with dimension-ordered routing.
It also implements a local bubble routing to avoid the ring deadlock.
This network can be fabricated with nearly equidistant wires by having consecutive logical tiles at a distance of two in the silicon.
A 32-bit 2D torus is 50\% bigger than a 2D mesh, but the torus provides twice the bisection bandwidth (BB) and 33\% fewer hops~\cite{ruche_batten}.

Since we target the design of \proj{} to be scalable even with hundreds of tiles per dimension (tens of thousands of tiles), scalability of network utilization is part of the design goal.
Scaling \proj{} up to the next power-of-two tiles per dimension results in four times the total tile count, but BB only doubles. While latency is greatly hidden by the pipeline effect, BB is critical for scalability since the vertex updates are irregular and result in communication to any tile.
This disparity increases the contention with network sizes.
To overcome that, we also explore ruche networks~\cite{ruche_batten,ruche_taylor}.

Ruche networks are long physical wires that bypass routers. 
They increase the router radix and decrease the latency from source to destination.
For example, in a network with a ruche factor of 2, a tile could route to its immediate neighbors or tiles at a distance of 2. 
A full ruche network of factor $R$ increases the BB by a factor of $(R-1)\times$ over the underlying network, so increasing sizes of $R$ with larger network sizes can compensate for the previously observed BB decrease.
Section~\ref{sec:results} provides a performance characterization of torus vs. mesh, with and without ruche networks.




\subsection{Scalable Memory Bandwidth}\label{sec:mem_bw_sec}
Large scratchpads are more area-efficient with modern, smaller, FinFET transistor nodes~\cite{dense_ff5nm,renesas_ff7nm, mediatek_sram_ff5nm}. 
This allows packing more SRAM on-chip than ever before. 
A real-world example is the Wafer-Scale integration of Cerebras, which enables 40GB of on-chip SRAM storage~\cite{cerebras_hotchips}.

A dedicated scratchpad memory per tile enables immediate, uncontested access to memory.
The width of each memory port is equal to the network width.
The PU can make one memory read and one write per cycle.
This is leveraged by instructions writing to queues from data arrays.

The PU has another port to fetch instructions from the scratchpad.
While the scratchpad is heavily banked to save per-access energy, not all banks need to have all ports, e.g., the instruction port can exist only for a fraction of the local memory, setting a limit to the code size.

Since \proj{}'s memory is distributed on-chip, the aggregated memory bandwidth increases linearly with the number of tiles, unlike many modern hardware architectures.

\section{Evaluation Methodology}\label{sec:eval}

In addition to four graph algorithms, we evaluated one sparse linear algebra kernel to demonstrate the generality of our approach for memory-bound applications.

We adapted the following \textbf{applications} from the GAP benchmark~\cite{gap_bench} and GraphIt~\cite{graphit}, splitting the program into tasks at each indirect memory access:
\textit{Breadth-First Search (BFS)} determines the number of hops from a root vertex to all vertices reachable from it;
\textit{Single-Source Shortest Path (SSSP)} finds the shortest path from the root to each reachable vertex;
\textit{PageRank} ranks websites based on the potential flow of users to each page~\cite{pagerank};
\textit{Weakly Connected Components (WCC)} finds and labels each set of vertices reachable from one to all others in at least one direction (implemented using graph coloring~\cite{connected_components});
\textit{Sparse Matrix-Vector Multiplication (SPMV)} multiplies a sparse matrix with a dense vector.

\textbf{Datasets:}
We used real-world networks and synthetic datasets. 
We use several different sizes of synthetic RMAT graphs~\cite{kron} of up to 67M vertices (V) and 1.3B edges (E), with up to 12GB of memory footprint, as well as real-world graphs: LiveJournal (V=5.3M, E=79M), Wikipedia (V=4.2M, E=101M) and Amazon (V=262K, E=1.2M).

\subsection{Simulation Approach}\label{sec:methodology}

We built a C++ cycle-level simulator to evaluate \proj{} due to the novelty of the execution model and the simplicity of the architecture.
Task instructions are modeled at each clock cycle; the network flits flow from tile to tile, from source to destination, one cycle per hop.
Our simulations are validated to provide correct program outputs over sequential x86 executions of the applications we evaluate~\cite{gap_bench}.

Current architectures pay an enormous energy cost associated with accessing off-chip memory~\cite{morc}.
Reading data locally consumes nearly two orders of magnitude less energy than moving an equivalent block of data across 15mm of on-chip wire~\cite{wire_energy} and three orders of magnitude less energy than bringing the data from off-chip DDR3~\cite{micron_dram}.

\textbf{Power and Area Model}:
With the latest technology, SRAM can achieve very high densities in FinFET, ranging from 29.2 Mb/$mm^2$ in 7nm~\cite{renesas_ff7nm} to 47.6 Mb/$mm^2$ in 5nm~\cite{dense_ff5nm}.
New technologies like FinFET or CMOS-ULVR show that leakage power can get as low as 1.7nW per cell~\cite{finfet}, 5$\mu$W per 8KB (8192 bytes) macro~\cite{ultralow-sram}, or 16.9$\mu$W for a 32KB macro at 7nm~\cite{renesas_ff7nm}.
Although 5nm is achievable today~\cite{mediatek_sram_ff5nm}, we will be using 7nm for our area and power models since this is a more mature transistor process, and we were able to find more reliable numbers for SRAM and logic.
Reading a bank of data consumes 5.8pJ, and writing it 9.1pJ, with an access time of 0.82 ns~\cite{renesas_ff7nm}.
Thus, we use a 1GHz frequency for \proj{} to fit the memory access time within the cycle length.

In addition to area-efficient SRAM technology, \proj{} uses processing units that resemble slim cores. 
We estimate the PU area considering the RISC-V Celerity, Snitch, and Ariane cores~\cite{celerity,snitch,ariane-riscv}.
To determine the dynamic and leakage power of the single-issue in-order core, we use the reports from Ariane~\cite{ariane_cost} and transistor power-scaling ratios to calculate the energy of those operations on a 7nm process~\cite{logic_7nm,finfet_7nm}.

Regarding the NoC, we explore a 2D-Torus and a 2D-mesh, with and without ruche networks.
These are modeled hop by hop between the routers, making the simulator precise for cycle count and energy.
We use 8pJ as the energy to move a 32-bit flit one millimeter\cite{piton_power} and assume the energy of moving a flit at the router to be similar to an ALU operation.
The NoC area is calculated based on Ou et al.~\cite{ruche_batten}.
These NoC options are explored in Fig.~\ref{fig:network_characterization}, and for other results \proj{} uses a regular torus NoC for grids up to 32$\times$32 (1024 tiles) and a torus NoC with ruche channels for larger grids.




\subsection{Comparison with the State-of-the-Art}\label{sec:pim_eval}

We looked at the literature on graph accelerators, i.e., those with a specialized hardware pipeline to process the different stages of graph traversing, and focused on Polygraph~\cite{polygraph}---the latest work.
We evaluated the code that the authors kindly provided and experimentally confirmed that Polygraph's performance plateaus with configurations larger than 16 cores while \proj{} continues to scale.
This is expected since that configuration already saturates the 512GB/s of DRAM bandwidth provided by their 8 memory controllers using High-Bandwidth Memory (HBM).

However, PIM-based proposals have a memory system that scales with the number of computing resources.
From these works, we compare ourselves with Tesseract~\cite{tesseract}.
We have simulated a more recent PIM-based approach, GraphQ~\cite{2019graphq}, which reported 3.3$\times$ performance and 1.8$\times$ energy improvements on average over Tesseract.
However, despite communicating with the authors and using their methodology, we were unable to obtain the cycle and energy values from simulation outputs that matched their results.
We do not include GraphQ in our discussion because we obtained 3.2$\times$ and 4.6$\times$ larger runtime and energy consumption (in geomean) than in \cite{2019graphq}.

We evaluate Tesseract using a Hybrid Memory Cube (HMC) configuration of 16 cores per cube (one per vault), aggregating a total of 256 cores among the 16 cubes.
To match their core count, we use a $16 \times 16$ \proj{} grid with 4.2MB of memory per tile.
To simulate Tesseract, we follow their methodology by using the Zsim simulator~\cite{sanchez2013zsim} with a 3D-memory power model~\cite{hmc_core_model}.
For the energy spent by the cores, we use the same power model as \proj{}---based on a 7nm transistor node.
We compare runtime and energy performance for the duration of the graph processing time, not considering loading the dataset from disk to HMC or to the \proj{} chip.
Although Tesseract already showed significant speedups over server-class OoO systems, we collected the cycle count of our x86 server while validating the output of graph applications to confirm that both Tesseract and \proj{} perform much better thanks to their scalable memory bandwidth (MBW).

\section{Results}\label{sec:results}

\subsection{Improvements over the HMC-based prior work}

This section demonstrates the gains in performance and energy efficiency of \proj{} over Tesseract~\cite{tesseract}, both using an equal number of cores (256).
We break down the \textbf{impact of the different optimizations of \proj{}} by starting with Tesseract and then evaluating \proj{} by adding one feature at a time to gradually reach full \proj{}.

\begin{figure}[htp]
\subf{\includegraphics[width=\columnwidth]{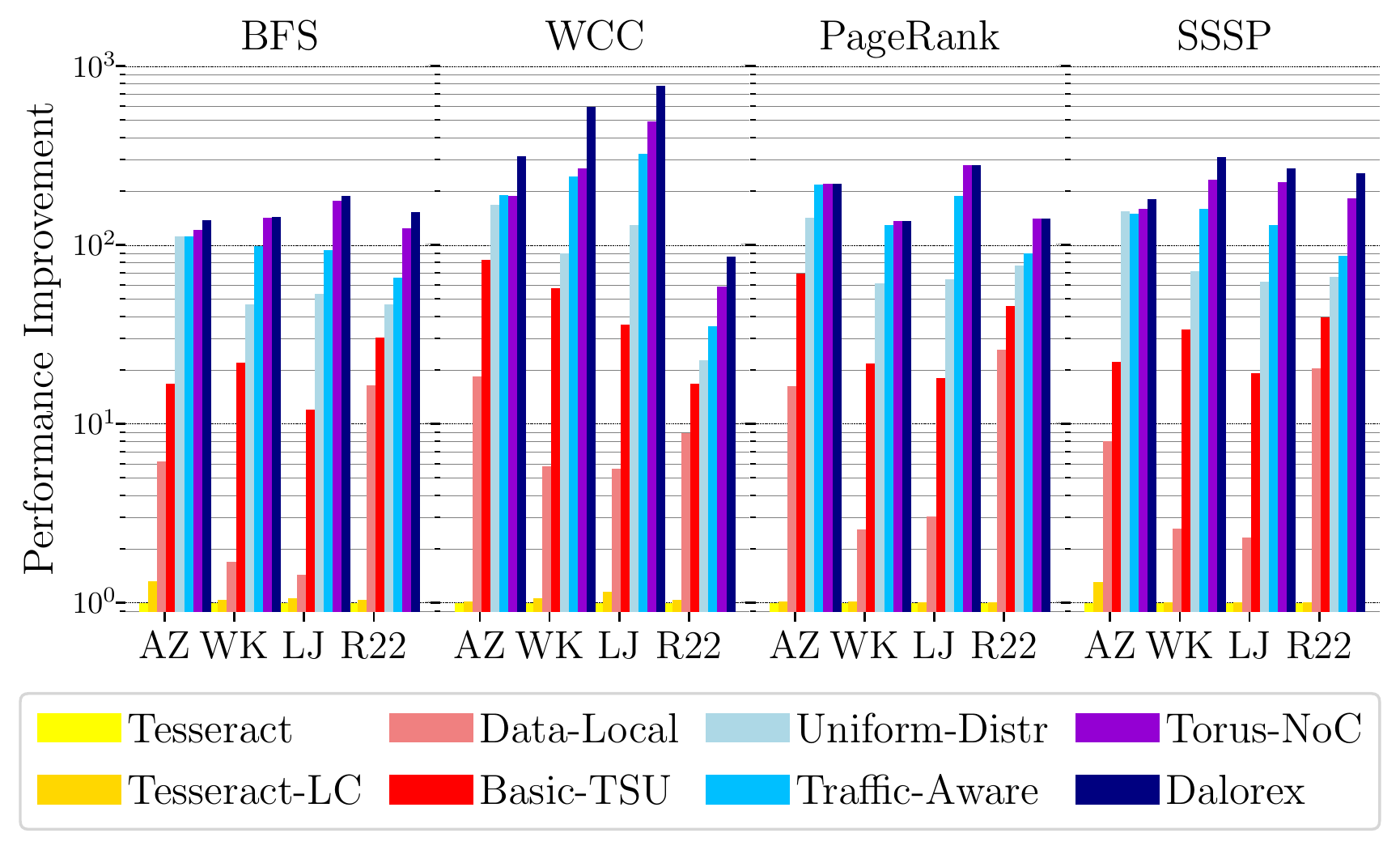} }
\\
\subf{\includegraphics[width=\columnwidth]{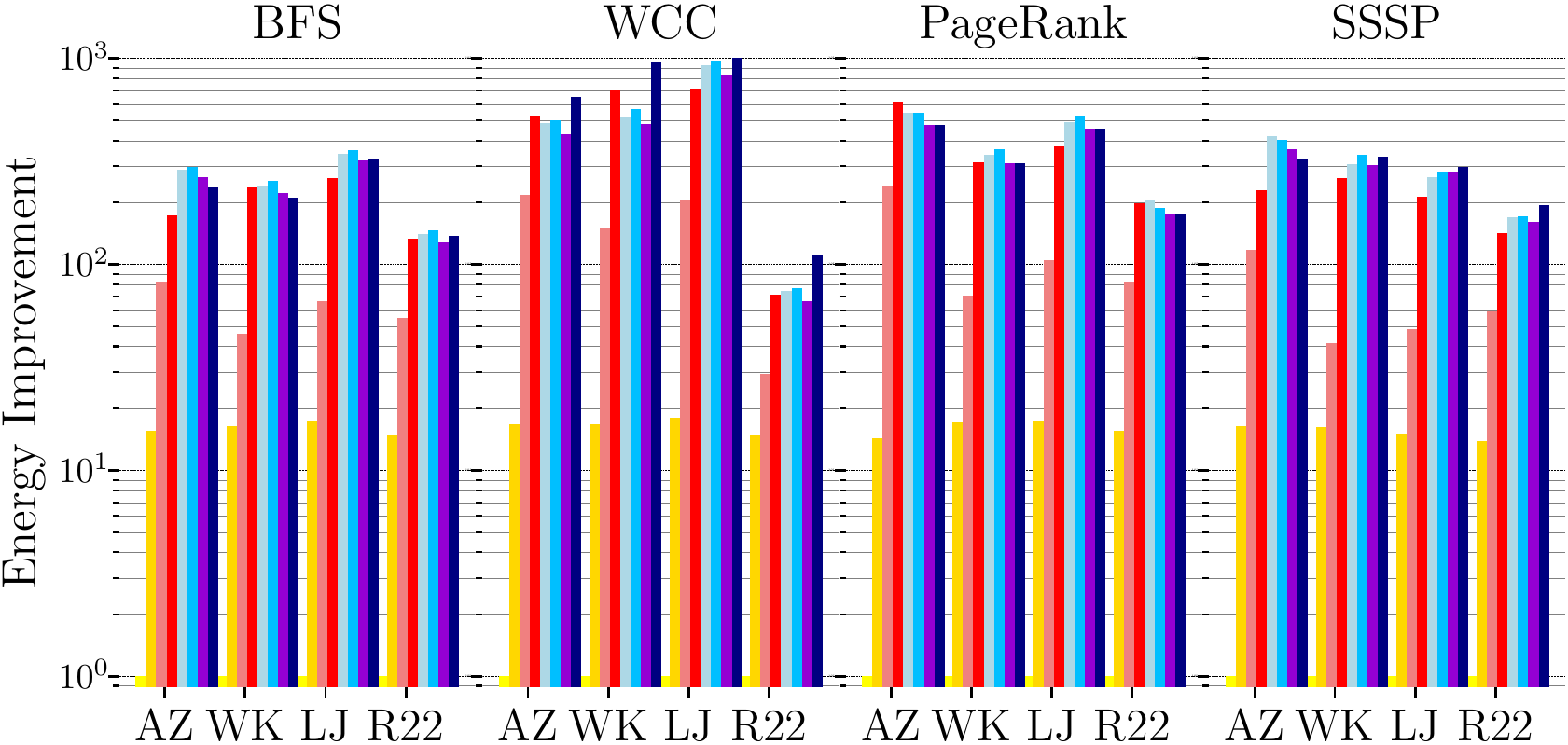} }
{}
\vspace{-3mm}
\caption{
Improvements in performance and energy efficiency achieved with the addition of large caches (LC) to Tesseract, and six \proj{} optimizations normalized to Tesseract.
All configurations use 256 processing cores.
X-axis studies four datasets.
Y-axis is logarithmic, higher is better.
Note that since PageRank necessitates per-epoch synchronization, the last datapoint (\proj{}-full) still uses a global barrier; thus, the bars do not change.
}
\vspace{-1mm}
\label{fig:pim_performance}
\end{figure}

We provision Tesseract with a 2MB private cache per core (512 MB of aggregated cache) and remove DRAM background energy, to approximate the impact of using distributed SRAM on performance and energy (\textit{$Tesseract-LC$}).
We evaluate \proj{} without \device{} by using the program flow of Tesseract (with remote vertex updates interrupting) to study the impact of vertex-based (Tesseract) versus \proj{}'s array chunking and task splitting (\textit{$Data-Local$}).
We add a basic \device{} to invoke tasks with round-robin scheduling to evaluate the impact of non-blocking and non-interrupting communication (\textit{Basic-\device{}}).
We distribute the data arrays by low-order bits instead of high-order bits to evaluate the impact of data placement. (\textit{$Uniform-distr$}). 
We add the scheduling policy that prioritizes tasks based on queue occupancy (\textit{$Traffic-aware$}).
We use a 2D Torus instead of a 2D Mesh (\textit{$Torus-NoC$}).
Finally, we removed the global barrier synchronization after each epoch, reaching $\proj{}-full$.
Section~\ref{sec:characterizations} further analyzes the impact of the NoC by testing one more design that benefits larger \proj{} grids.

\textbf{\textit{Performance:} }
Fig.~\ref{fig:pim_performance} (top) shows that \proj{} substantially outperforms Tesseract across all datasets and applications.
The dataset-chunking strategy and the data-local execution on \proj{}'s homogeneous architecture improves performance by 6.2$\times$ (\textit{$Data-Local$}).
On top of that, non-blocking and non-interrupting task invocation through \device{} improve performance by 4.7$\times$,
uniform data placement improves another 2.6$\times$, and traffic-aware scheduling 1.7$\times$ more.
Finally, removing the barriers and upgrading the NoC provides an extra 1.8$\times$, compounding a 221$\times$ geomean improvement.
Synchronization in graph workloads causes each epoch to take as long as the slowest tile's execution, increasing the total runtime.
WCC, having more epochs, benefits the most from barrierless processing.
Similarly, Wikipedia (WK) benefits most from removing the barrier as its graph structure leads to more epochs.

\textbf{\textit{Energy:}}
Fig.~\ref{fig:pim_performance} (bottom) shows that the performance improvement of \proj{} over Tesseract, together with its power-efficient design, yields a very large improvement in energy consumption.
The compound improvements from using SRAM technology (16$\times$), our architecture and data layout (5.2$\times$), and \device{} (3.9$\times$), total a geomean of 325$\times$.
The breakdown shows that the energy of refreshing DRAM has the biggest impact on Tesseract (also concluded in their paper).
Upgrading the NoC to a 2D Torus increases energy usage by 12\% geomean, besides being 40\% faster.
This is because the Torus network is more power-hungry and has longer wires.

\textbf{\textit{Area:}}
The $16 \times 16$ \proj{} with a 4.2MB memory per tile uses much less chip area (305$mm^2$) than the aggregated area of the 16 cubes of Tesseract (3616$mm^2$).
The prior work based on HMC is constrained by the number of cores that can be integrated into the logic die of a memory cube---often one per vault.
Since a cube contains 8GB, a core accesses a 512MB DRAM vault.
Most of this DRAM space is unused with the datasets that Tesseract evaluated (the empty bitlines of DRAM are switched off to save power).
However, larger datasets would have a much bigger runtime in the HMC-based design since it cannot have more cores without increasing the number of memory cubes.

HMC is also limited in scalability due to the large \textit{power density}, which makes it more challenging to cool in 3D integrations~\cite{eckert2014thermal}.
In \proj{}, power is evenly distributed, so power density stays below 300mW/$mm^2$ for all our experiments.
This is much below the limits of air-cooled 2D chips ($\sim$1.5W/$mm^2$~\cite{power_density}).
Another limitation of the HMC-based approaches is that lower inter-cube \textit{communication bandwidth} made their authors consider graph partitioning per cube, limiting the scalability of each partition.

To summarize, such large across-the-board improvements are achieved with a conjunction of optimizations at different levels of the software-hardware stack. \proj{}:
(1)~reduces data movement with its \textit{fully} data-local program execution model, where only task parameters are moved;
(2)~task invocations are natively supported through the \device{}, so there are no interrupt overheads as in prior remote procedure calls;
(3)~task are scheduled based on the network's task traffic;
(4)~does not require barrier synchronization between tiles for BFS, WCC, and SSSP;
(5)~decouples the placement of vertices and edges, giving an equal number of edges to each tile, improving work balance;
(6)~utilizes SRAM to store the tile-distributed dataset, making data access immediate and energy-efficient for the challenging fine-grain irregular accesses. 
Also, SRAM is offered in finer sizes than DRAM.
This enables \proj{} to use a few megabytes of memory per tile, which we found to be energy-optimal in the scaling experiments of the next section.

\subsection{\proj{} Scales Beyond Thousands of Tiles}\label{sec:scalability}

\begin{figure}[htp]
\vspace{-3mm}
\subf{\includegraphics[clip,width=\columnwidth]{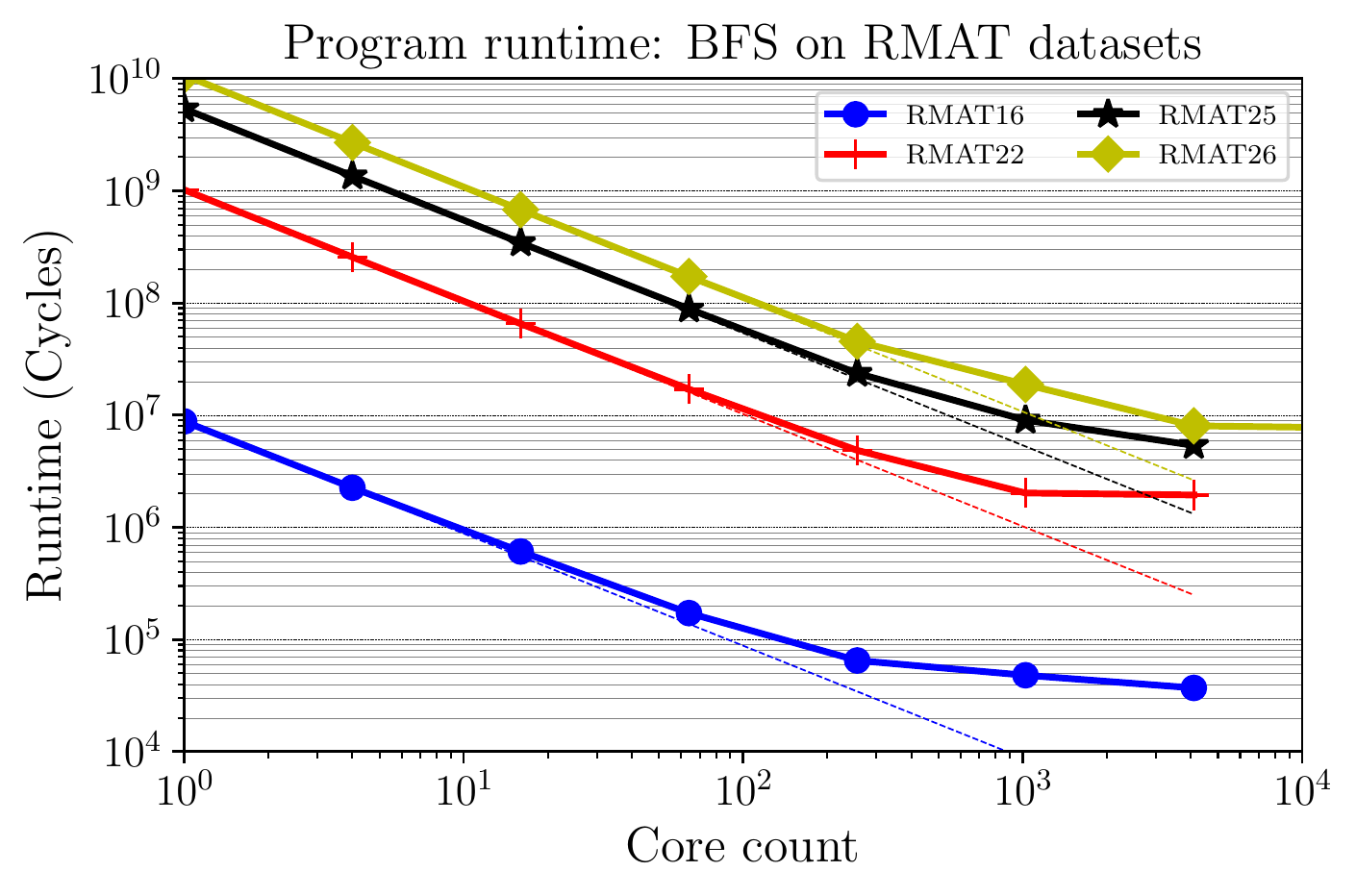} }
{}
\\
\subf{\includegraphics[clip,width=\columnwidth]{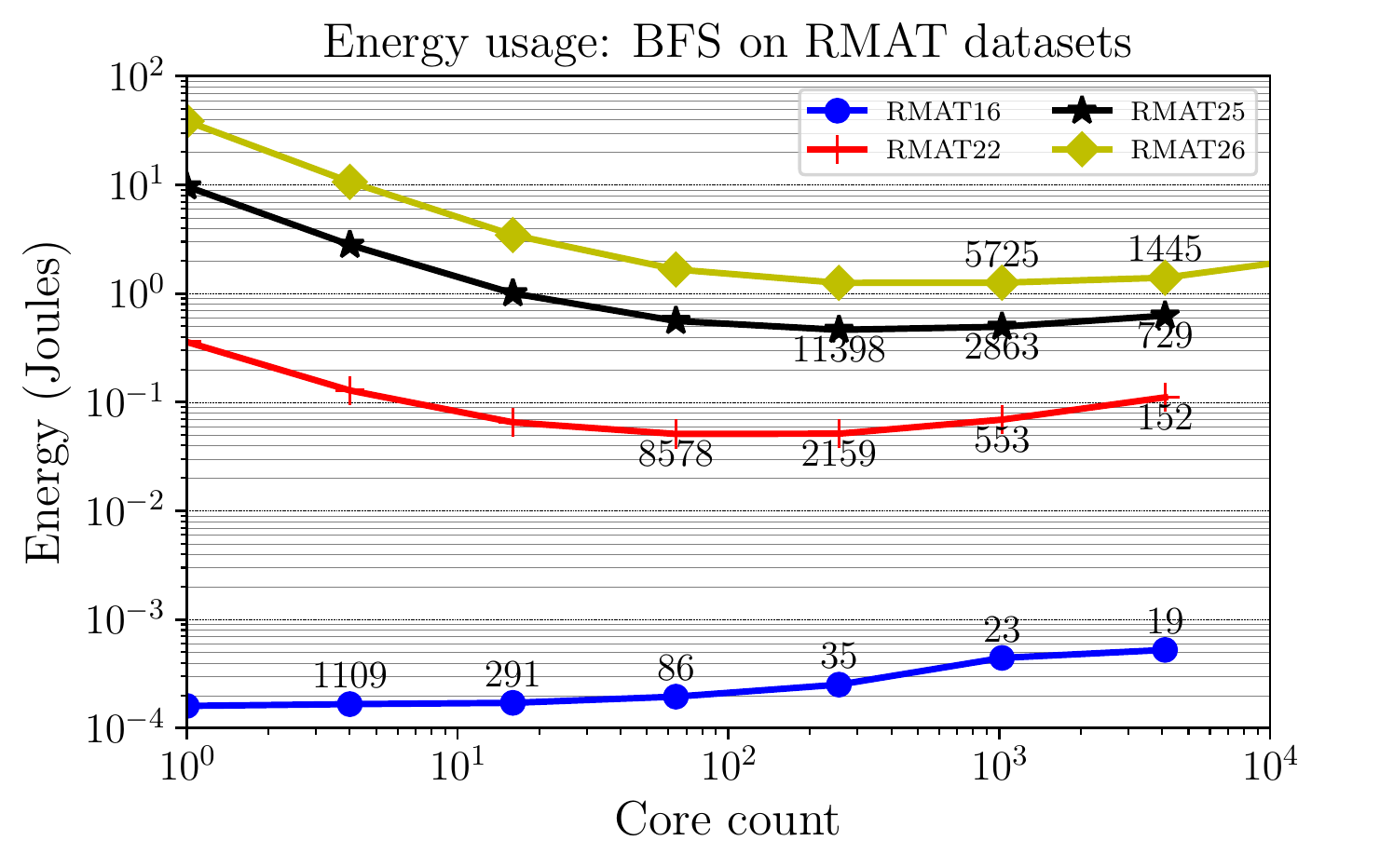} }
{}
\vspace{-6mm}
\caption{Analysis of the runtime (cycles) and energy consumption (Joules) of BFS for four RMAT datasets and scaling core counts.
}
\vspace{-1mm}
\label{fig:scaling_perf}
\end{figure}

Fig.~\ref{fig:scaling_perf} evaluates strong scaling for BFS with increasing sizes of \proj{} grids, and four RMAT datasets of size $2^{16}$, $2^{22}$, $2^{25}$ and $2^{26}$ vertices (average ten edges per vertex).

\textbf{\textit{Performance Scaling:}}
The upper plot of Fig.~\ref{fig:scaling_perf} shows the runtime (in cycles) of executing each dataset on \proj{} with an increasing number of tiles by multiples of four.
The most exciting result of this analysis is that \proj{} has a close to linear scaling until it hits the parallelization limit.
This occurs when the chunk of data per tile is $\sim$1,000 vertices, indicating that performance is not limited by MBW (as is the case for all previous work) but tiles starving for work.

\textbf{\textit{Energy Scaling:}}
To understand the optimal scratchpad size for energy consumption, we analyzed energy as the number of tiles increases and the required scratchpad size decreases.
Each scratchpad stores the dataset chunk, the program binary, and the queues.
Fig.~\ref{fig:scaling_perf} (bottom) shows the total energy spent to run BFS for different  numbers of tiles.
Next to each datapoint is the memory used by each tile (in KB).

As the tile count increases and the dataset chunk per tile becomes smaller, energy first decreases to a minimum and then increases.
This is mainly driven by leakage power.
Since the aggregated memory capacity remains nearly constant in this experiment, the SRAM leakage power becomes less significant with larger parallelizations.
Total energy keeps decreasing while the cores are fully utilized.
The deflection point, i.e., minimal energy execution for each dataset, arrives when the amount of data per tile is $\sim$10,000 vertices.
This optimal range is invariant with dataset size.
The smallest dataset (RMAT-16) reaches this point very early, being already past that point with 64 tiles, where each tile holds $\sim$1,000 vertices.

Fig.~\ref{fig:scaling_perf} shows that the performance keeps scaling beyond the energy-optimal configuration.
Unlike clusters, where performance degrades with small messages between computing nodes, \proj{} communication is at a word granularity.
Scaling is not limited by MBW but only when the workload per tile is so low that we lose the pipeline effect and the PUs starve for tasks.
While approaching the parallelization limit of a dataset, energy increases because of the leakage energy of the additional PUs that are not fully utilized.

\begin{figure}[t]
\centering  
\includegraphics[width=\columnwidth]{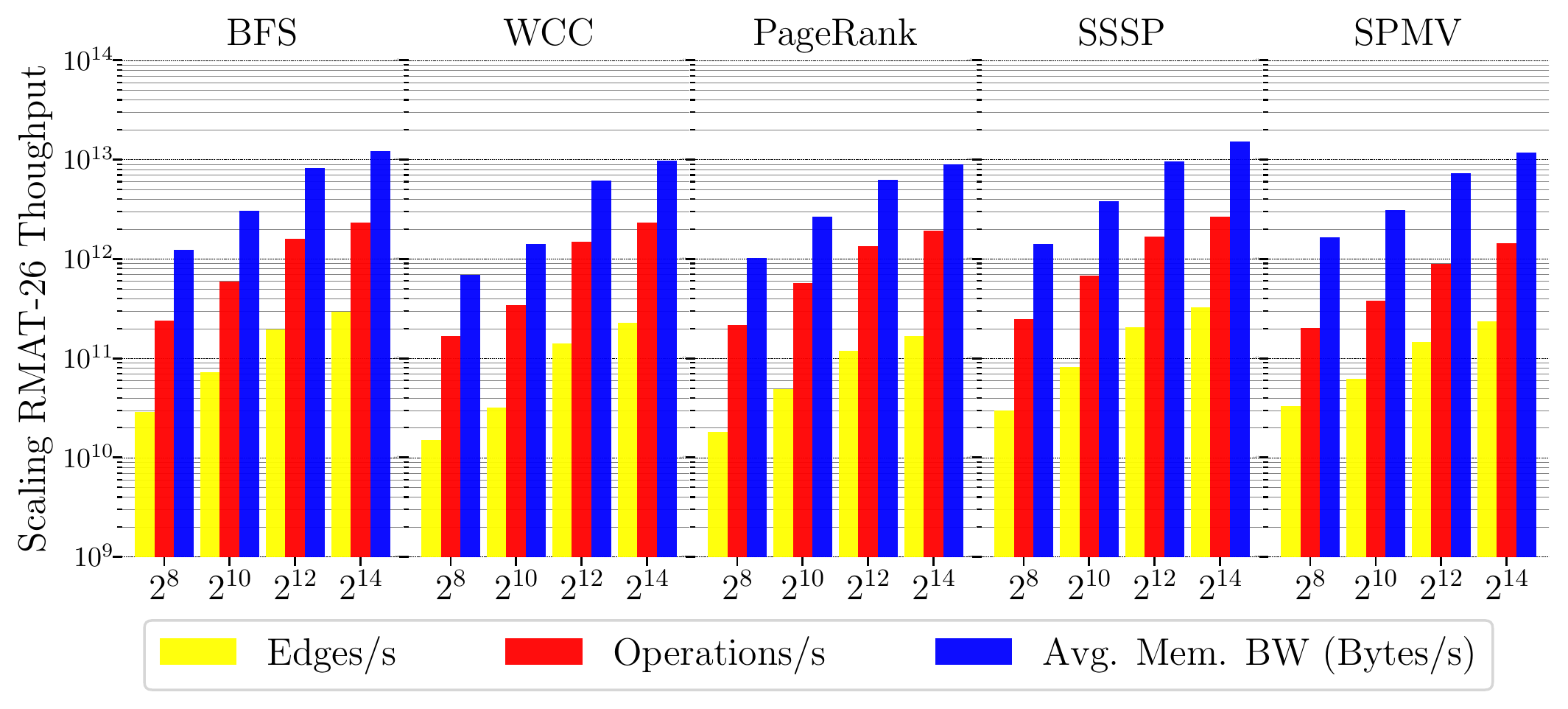}
\vspace{-5mm}
\caption{
Throughput in terms of executed instructions and edges per second, and the average on-chip memory bandwidth (MBW) used to achieve that.
The X-axis is the size of the \proj{} grid used when analyzing strong scaling RMAT-26, ranging from 256 to 16,384 tiles.
The Y-axis is logarithmic.
}
\vspace{-4mm}
\label{fig:scaling_bw}
\end{figure}

\textbf{\textit{Throughput:}}
Fig.~\ref{fig:scaling_bw} shows the number of edges and operations processed per second as a measure of the throughput and the average aggregated MBW by all tiles.
We study this with doubling counts on X- and Y-dimensions for all benchmarks running our biggest RMAT dataset.
We observe that both throughput and utilized MBW grow until the largest configuration we simulated: $128\times128$ ($2^{14}$, i.e., 16,384 tiles).
This last configuration reaches two teraoperations/s, using over ten terabytes/s of MBW.
The throughput and MBW of \proj{} for graph applications are beyond the reach of prior work in hardware accelerators using HBM and the 3D-memory integration that Tesseract proposed.
\proj{} MBW is possible by having all the memory storage distributed across tiles and linearly increasing the number of memory ports with the tile count.
The fact that the MBW does not saturate allows the PUs to process vertices and edges at the very high rates shown in Fig.~\ref{fig:scaling_bw}.
Note that we reported the average MBW of the whole program and not the maximum utilized at a given time.
The peak MBW available only increases with the number of tiles, which is 131TB/s on a $128\times128$ \proj{}.

\subsection{Characterizing the Network-on-Chip (NoC)}\label{sec:characterizations}

\begin{figure}[t]
\centering  
\includegraphics[width=\columnwidth]{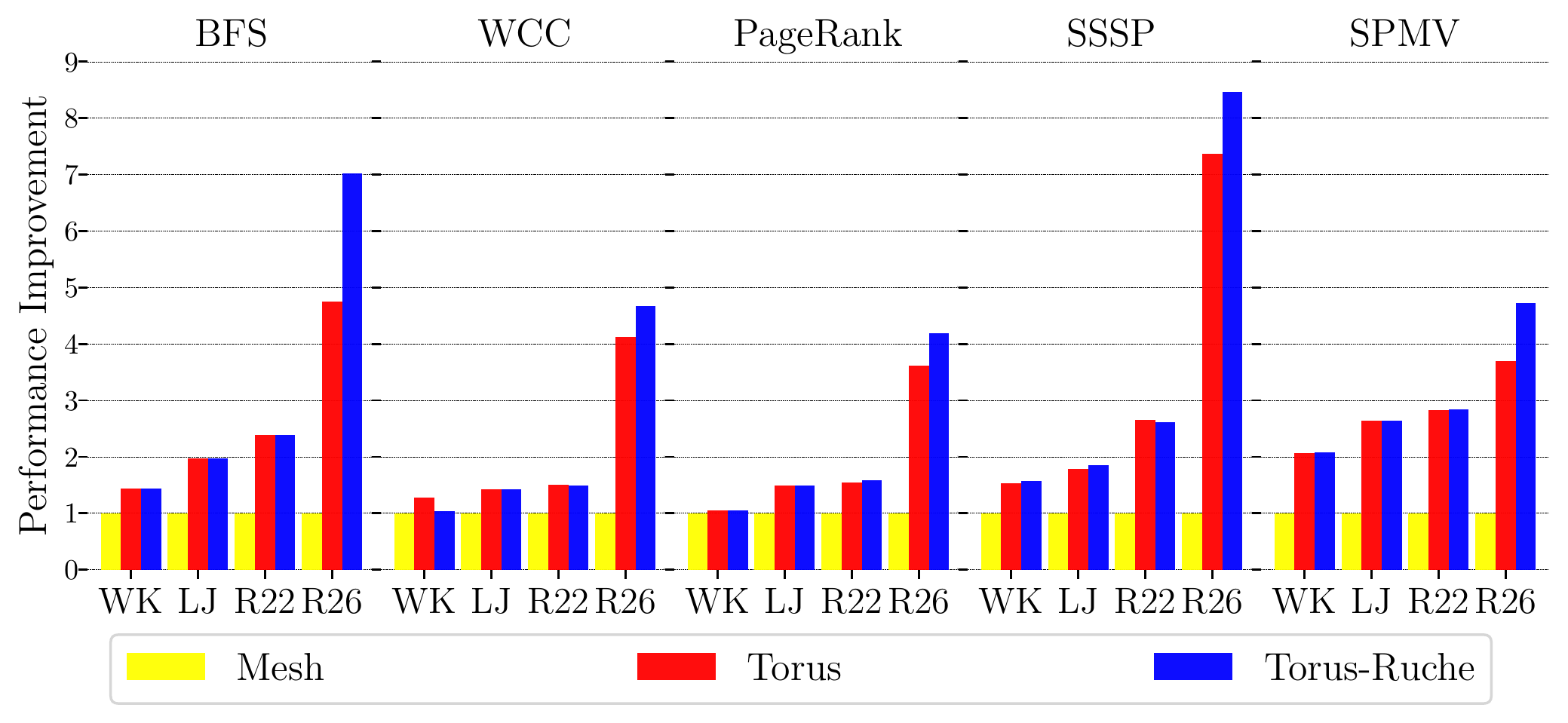}
\vspace{-5mm}
\caption{Performance improvement of Torus and Torus-Tuche over Mesh.
The X-axis shows the datasets used for each of the applications evaluated.
RMAT-26 runs on 64$\times$64 tiles, and the rest 16$\times$16.
}
\vspace{-3mm}
\label{fig:network_characterization}
\end{figure}

\begin{figure}[t]
\centering  
\includegraphics[width=\columnwidth]{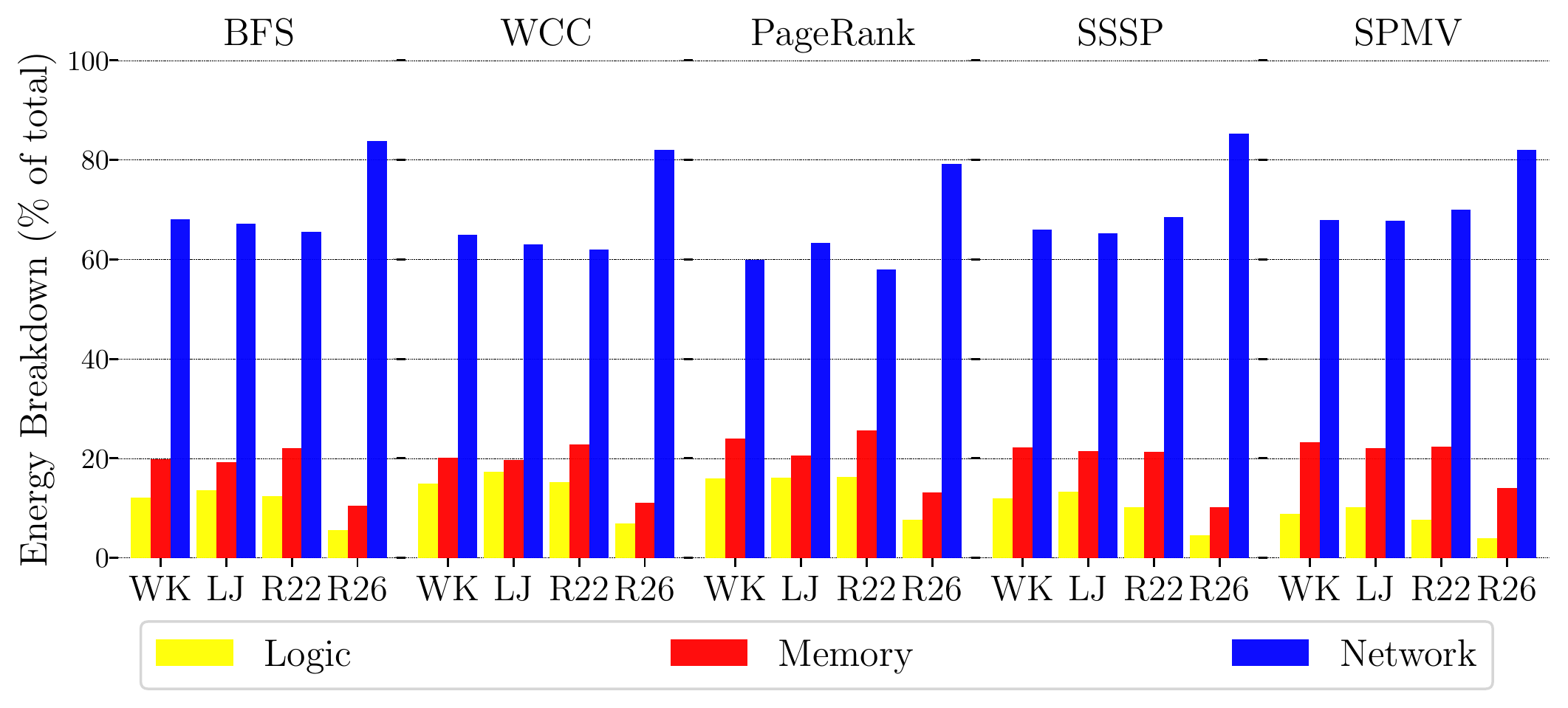}
\vspace{-4mm}
\caption{
Breakdown of the energy consumed by the computing logic, the SRAM cells, and the network communication (including routing and wire energy).
The Y-axis is the percentage of total energy spent running the program.
The X-axis shows the datasets used for each of the applications evaluated.
RMAT-26 runs on 64$\times$64 tiles, and the rest 16$\times$16.
}
\vspace{-4mm}
\label{fig:energy_breakdown}
\end{figure}

We analyze \proj{} with the three NoC types we considered in Section~\ref{sec:network_sec}: 2D mesh, 2D torus, and combining the torus with ruche channels.
Fig.~\ref{fig:network_characterization} shows the performance improvement of both torus options over the mesh.
Using a $16\times16$ torus is nearly twice as fast as a mesh, on average, for the smaller datasets (Wikipedia, LiveJournal, and RMAT-22), which justifies the area cost of an additional 0.2\% of the total chip area (using 4MB tiles).

\begin{figure}
\centering
\resizebox{\columnwidth}{!}{
\begin{tabular}{c c}
\subf{\includegraphics[width=40mm]{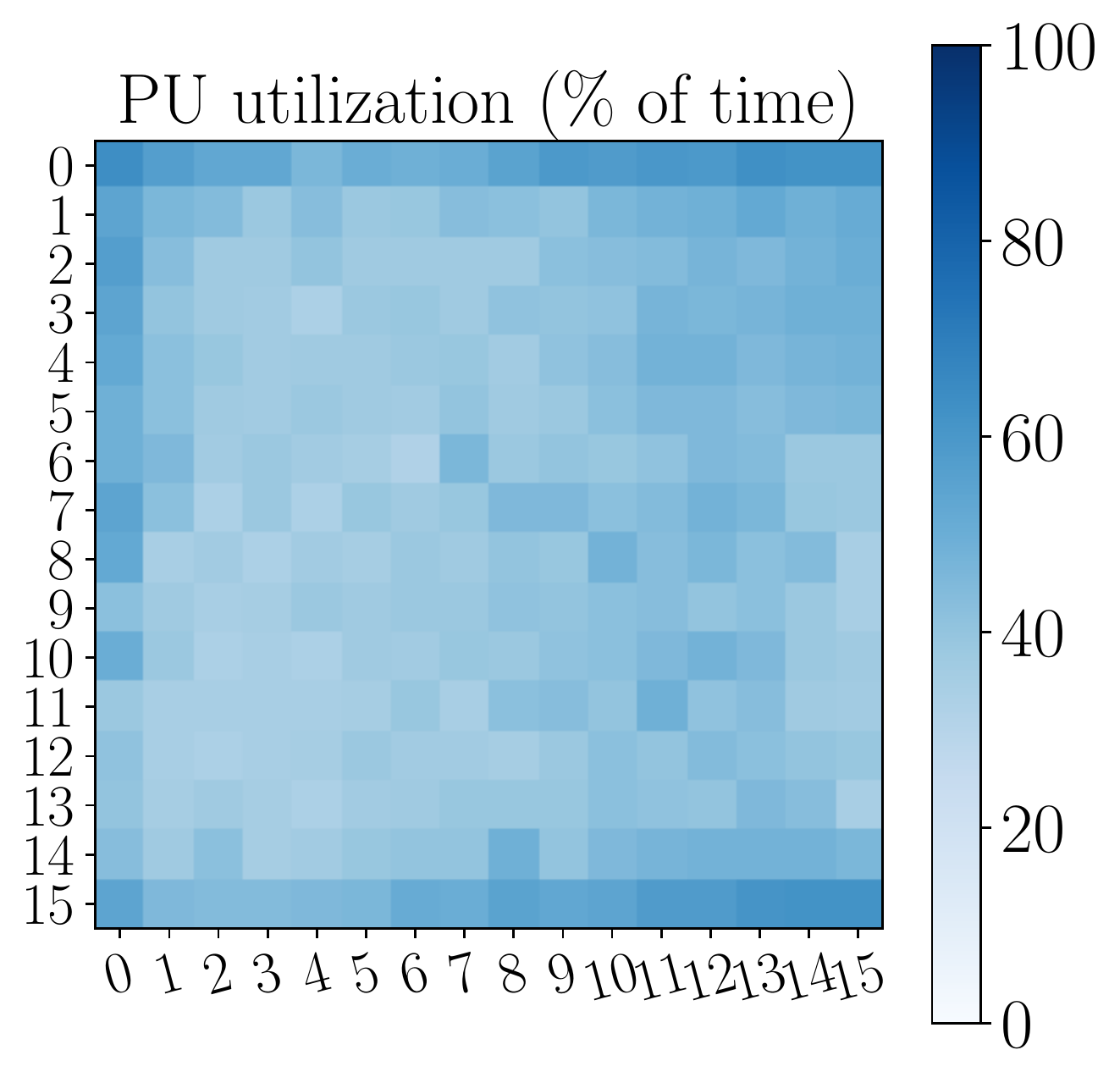}}
{(a) 2D Mesh}
&
\subf{\includegraphics[width=40mm]{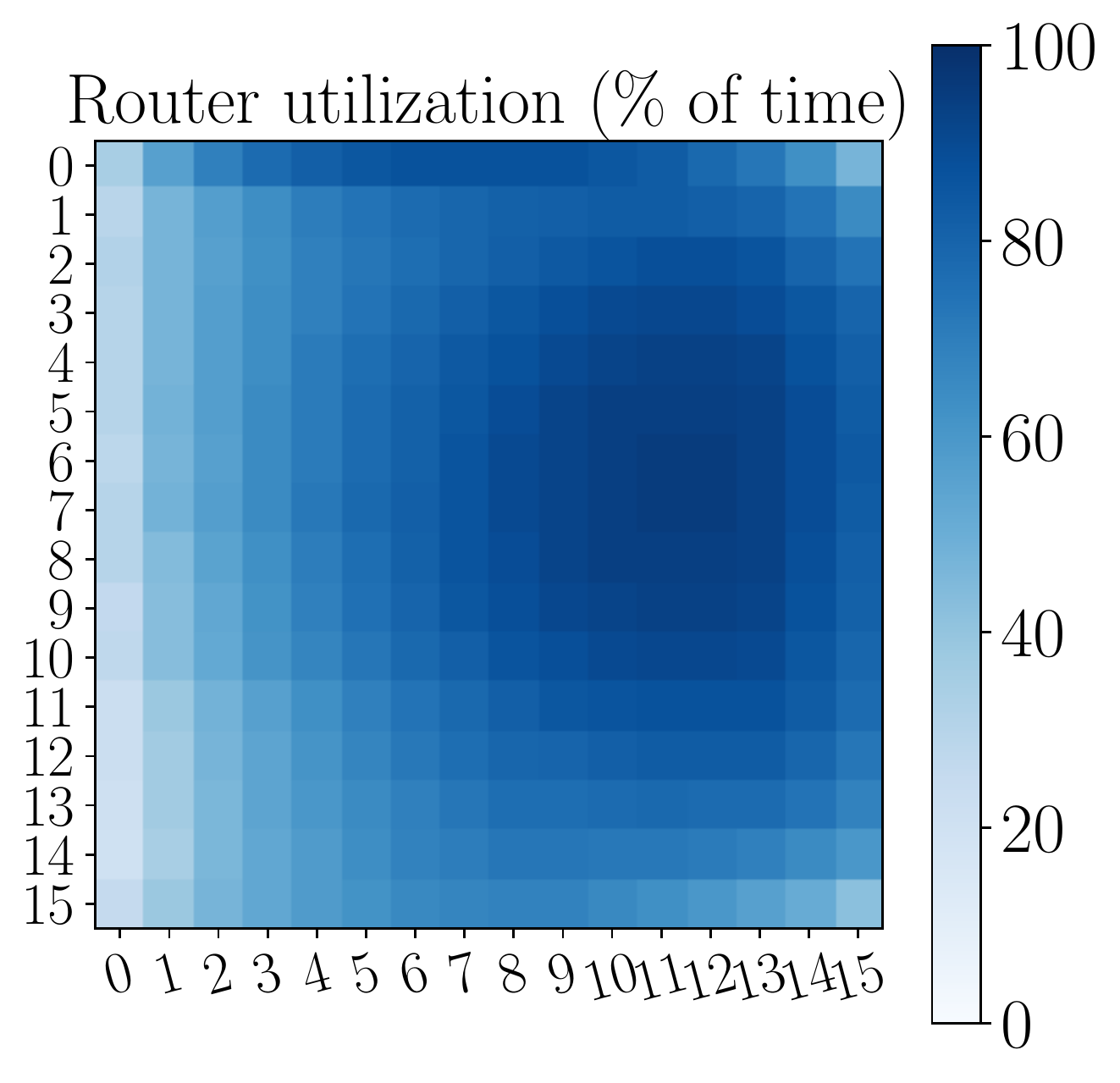}}
{}
\\
\subf{\includegraphics[width=40mm]{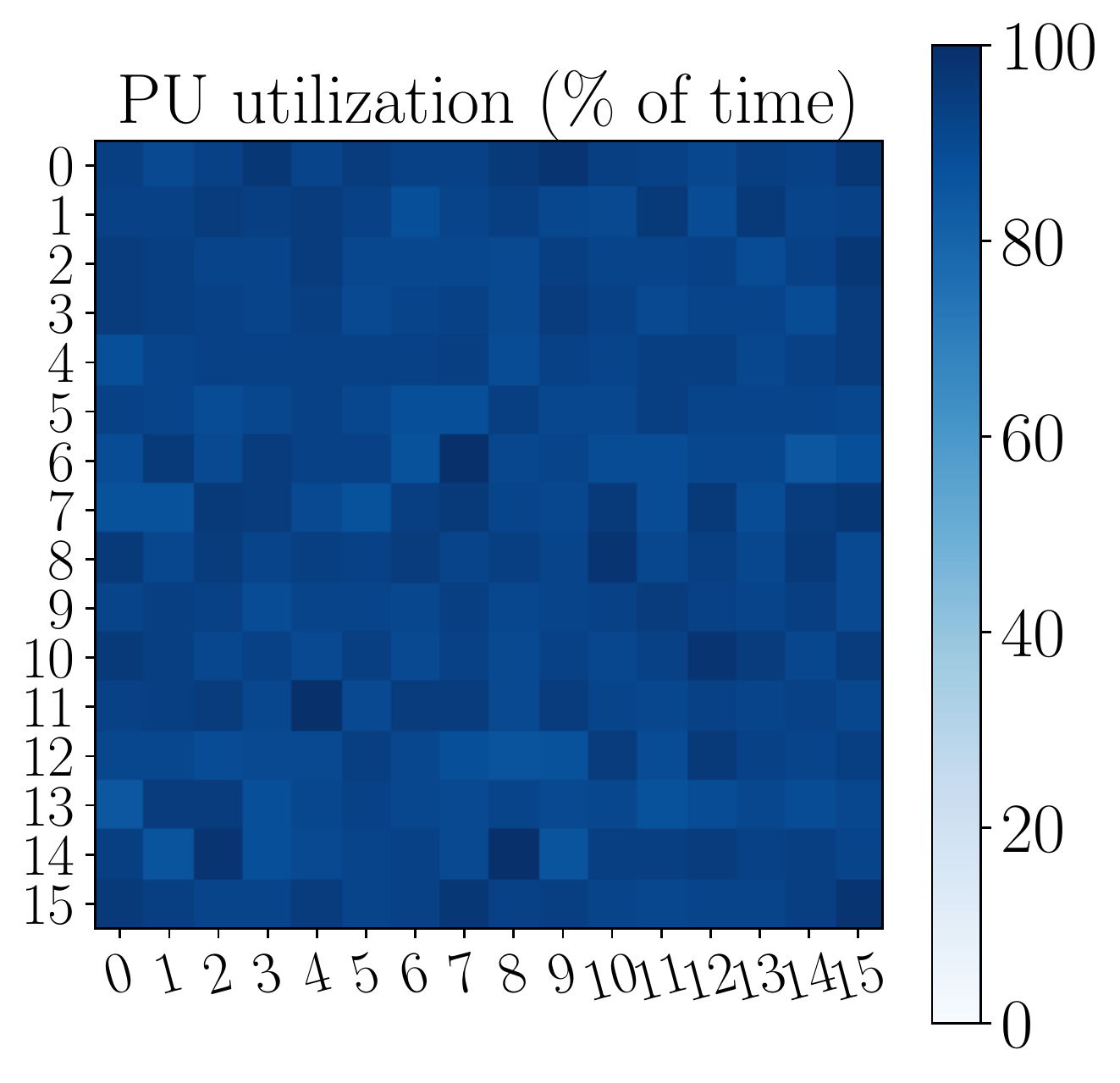}}
{(b) 2D Torus network}
&
\subf{\includegraphics[width=40mm]{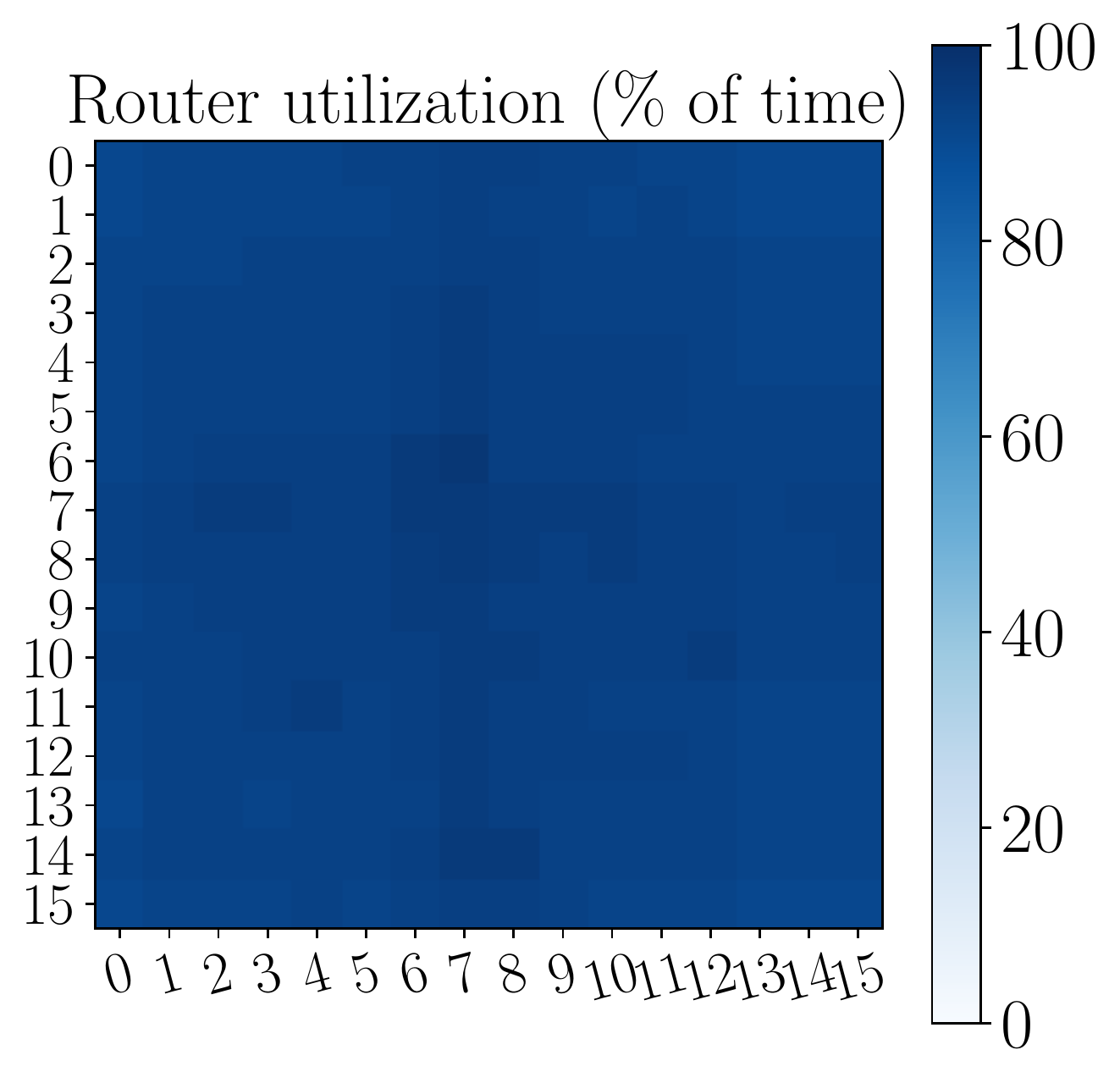}}
{}
\end{tabular}
}
\caption{Heatmaps of the utilization of PUs and routers as a percentage of runtime while running SSSP on RMAT-22.
The 16$\times$16 tiles are connected by a mesh (top) or torus (bottom).
}
\vspace{-2mm}
\label{fig:heatmaps}
\end{figure}

For this $16\times16$ grid, we generated heatmaps of the utilization of PUs and routers to visually 
demonstrate the advantage of torus over mesh.
Fig.~\ref{fig:heatmaps} shows that the contention towards the center of the mesh (top) clogs the NoC and makes the PUs starve for tasks, whereas the torus (bottom) has a uniform router utilization, unleashing the full potential of the PUs.

As shown in Fig.~\ref{fig:network_characterization}, the improvement of the torus NoC becomes even greater for the $64\times64$ grid used to evaluate RMAT-26.
We found \textit{ruche with torus} to only improve performance on the large grid.
Despite that the ruche-torus NoC uses more than twice the area of a regular torus (extra cost of 1.2\% area) and higher power, the performance gains at the $64\times64$ grid justify its use.
Although not shown here, we have found ruche combined with mesh not to be as effective as torus alone, despite its larger area cost.

Fig.~\ref{fig:energy_breakdown} breaks down the energy consumed by \proj{}.
We use a $16\times16$ grid to run WK, LJ, and RMAT-22, and $64\times64$ to run RMAT-26.
In \proj{}, the network consumes the most energy: the larger the network, the longer the average distance traveled to update a vertex, therefore, a greater share of the total energy consumption.
This is expected because \proj{} uses energy-efficient memories and very simple processing units (PU).
PUs are also not actively waiting for messages to arrive but are powered off by the \device{} when idle.

\section{Further related work \& Discussion}\label{sec:related}


Due to storage limitations, graph networks with trillions of edges inevitably need to be partitioned and processed by multiple systems.
However, within each system (where each partition is processed), distributed graph processing frameworks~\cite{giraph, giraph++, gemini, pregel} are bottlenecked by the memory hierarchies in current computer architectures.

\proj{}, and other HMC-based approaches for graph processing~\cite{2018graphp,2019graphq,tesseract} behave as a large accelerator for the host CPU.
Loosely-coupled accelerators for graph applications have been investigated before, e.g., ~\cite{graphicionado,ozdal,polygraph}.
However, ASIC accelerators cannot execute the variety of workloads that ISA-programmable processors can.
\proj{} is not restricted to executing graph applications and can be used for other domains since it is software programmable.

Unlike the prior work, which assumes an HMC technology that has not been adopted, there is already an example of an architecture where all the memory is distributed on a chip that is being commercialized by Cerebras~\cite{cerebras,cerebras_hotchips} to speed up Machine Learning applications.

As we show in this paper, the \proj{} execution model and design decisions like network type, local-memory size, and task scheduling unleash an outstanding graph processing performance for distributed-memory architectures.

Another example of an architecture that could use \proj{} is a regular memory hierarchy with large L1 caches per tile, leveraging prior work in cache-scratchpad duality~\cite{esp_cache,amoeba,cameo} to provide a hybrid solution that would benefit both cache-averse and cache-friendly workloads.







\section{Conclusion}\label{sec:conclusions}


%

\proj{} is designed to massively parallelize applications that are memory-bound due to the irregular memory accesses, for which prior work does not exhibit good strong scaling.
Datasets larger than the aggregated memory capacity of a \proj{} chip could be partitioned across several chips, having each subproblem parallelized to the extreme.

Since \proj{} is ISA-programmable, it is applicable to any application fitting in the task-based programming model, although it is most advantageous for those bottlenecked by pointer indirection or atomic operations, e.g., SPMV.

This paper demonstrates strong scaling on four graph algorithms and one sparse linear algebra kernel.
We found that reaching the parallelization limits requires:
~(1) a \textit{uniform work balance}, which we achieve with an equal amount of data per tile;
~(2) architectural and programming \textit{support to invoke remote tasks natively} with no message overhead;
~(3) a \textit{traffic-aware} arbitration of tasks;
~(4) a \textit{NoC that minimizes contention}, where a torus is superior to the mesh; and
~(5) a \textit{NoC that scales bisection bandwidth} with very large tile counts on a 2D silicon by adding ruche channels to the torus.
Together with the data partitioning scheme that allows barrierless frontiers, these optimizations make \proj{} two orders of magnitude faster and more energy-efficient than the state-of-the-art in PIM-based graph processing.






\section*{Acknowledgments}
This material is based on research sponsored by the Air Force Research Laboratory (AFRL), Defense Advanced Research Projects Agency (DARPA) under agreement FA8650-18-2-7862, and National Science Foundation (NSF)
award No. 1763838.~\footnote{The U.S. Government is authorized to reproduce and distribute reprints for Governmental purposes notwithstanding any copyright notation thereon. The views and conclusions contained herein are those of the authors and should not be interpreted as necessarily representing the official policies or endorsements, either expressed or implied, of NSF, AFRL and DARPA or the U.S. Government.}
We thank Tyler Sorensen for his useful feedback.

\balance
\bibliographystyle{IEEEtranS}
\bibliography{refs}

\end{document}